\documentclass{nature}
\usepackage{amssymb}
\usepackage{amsmath}
\usepackage{changes}

\usepackage{graphicx}
\usepackage{color}
\usepackage[normalem]{ulem} 
\usepackage{soul}
\usepackage{url}
\usepackage[colorlinks]{hyperref}
\usepackage{ulem}
\usepackage{mathptmx}

\def\deg{\hbox{$^\circ$}}

\usepackage{subfig}
\usepackage[font=footnotesize]{caption}
\RequirePackage{lineno}
\usepackage{multibib}
\newcites{Main}{References}
\newcites{Method}{Methods References}

\DeclareRobustCommand{\ion}[2]{%
\relax\ifmmode
\ifx\testbx\f
{\mathbf{#1\,\mathsc{#2}}}\else
{\mathrm{#1\,\mathsc{#2}}}\fi
\else\textup{#1\,{\textsc{#2}}}%
\fi}
\DeclareMathAlphabet{\mathsc}{\encodingdefault}{\familydefault}{m}{sc}

\newcommand{\araa}{Annu. Rev. Astron. Astrophys.}   
\newcommand{\aj}{Astron. J.}   
\newcommand{\apj}{Astrophys. J.}   
\newcommand{\apjl}{Astrophys. J. Lett.}   
\newcommand{\apjs}{Astrophys. J. Suppl. Ser.}   
\newcommand{\apss}{Astrophys. Space Sci.}   
\newcommand{\aap}{Astron. Astrophys.}   
\newcommand{\mnras}{Mon. Not. R. Astron. Soc.}   
\newcommand{\prd}{Phys. Rev. D}   
\newcommand{\ssr}{Space Sci. Rev.}   

\usepackage{floatrow}
\DeclareFloatFont{tiny}{\scriptsize}
\title{Effective Shielding of  $\lesssim$ 10 GeV Cosmic Rays from Dense Molecular  Clumps}

\begin{document}

\maketitle
\author{Rui-zhi Yang$^{1,2,3}$,
Guang-Xing Li$^{4}$,
Emma de O\~na Wilhelmi$^{5}$,
Yu-Dong Cui$^{6}$,
Bing Liu$^{1,2,3}$,
Felix~Aharonian$^{7,8,9}$
}

\begin{affiliations}
\small
\item Deep Space  Exploration Laboratory/School of Physical Sciences, University of Science and Technology of China, Hefei 230026, China
\item CAS Key Laboratory for Research in Galaxies and Cosmology, Department of Astronomy, University of Science and Technology of China, Hefei, Anhui 230026, China
\item School of Astronomy and Space Science, University of Science and Technology of China, Hefei, Anhui 230026, China
 \item South-Western Institute for Astronomy Research (SWIFAR), Yunnan University (YNU), Kunming 650500, People’s Republic of China
\item Deutsches Elektronen Synchrotron DESY, 15738 Zeuthen, Germany
\item School of Physics and Astronomy, Sun Yat-Sen University, Guangzhou, 510275, China
\item Dublin Institute for Advanced Studies, 31 Fitzwilliam Place, Dublin 2, Ireland 
\item Max-Planck-Institut f\"ur Kernphysik, P.O. Box 103980, D 69029 Heidelberg, Germany 
\item Gran Sasso Science Institute, 7 viale Francesco Crispi, 67100 L'Aquila,  Italy
\end{affiliations}

\hfill

\begin{abstract}
The density of cosmic rays inside molecular clouds determines the ionization rate in the dense cores where stars form. It is also one of the drivers of astrochemistry leading to the creation of complex molecules. Through Fermi Large Area Telescope observations of nearby giant molecular clouds, we observed deficits (holes) in the gamma-ray residual map when modelling with the expected gamma-ray diffuse emission from uniform cosmic rays interacting with the molecular content. We propose that the deficit is due to the lack of penetration of the low-energy (sub-GeV to GeV) cosmic rays into denser regions or clumps. This differs from the prevailing view of fast cosmic ray transport in giant molecular clouds where the magnetic turbulence is suppressed by neutral-ion damping, as our results require a slow diffusion inside dense molecular clumps. Through modelling we find that while the shielding is negligible on the cloud scale, it becomes important in the denser, parsec-sized regions where the gravitational collapse is already at play, changing the initial condition of star formation and astrochemistry.
 
 \end{abstract}
\newpage

 Cosmic rays (CRs) are one of the most important ingredients in the interstellar medium (ISM), with an energy density similar to that of the magnetic field and the thermal energy of the gas. The interplay between star formation and CRs is one of the most important physical processes in astrophysics. The star formation process can be regarded as the ultimate energy source of CRs, as the stars either accelerate CRs through the stellar winds\cite{aharonian19} or via supernova remnants and compact objects at later stages of their evolution, whereas the CRs generated by star formation can change the initial condition of star formation as well as the astrochemistry processes therein.

Star formation occurs in giant molecular clouds (GMCs). These massive cloud-like entities are made of cold, molecular gas.
CRs also play an essential role in star formation. CRs govern the heating and ionization processes in the
star-forming regions \cite{dalgarno06}, affect the dynamical evolution of gas,
and initiate several crucial chemical reactions in the dense cores of molecular
clouds \cite{papadopoulos10}. The CR density may even affect the outcome of the
star formation measured in terms of the initial mass function\cite{papadopoulos10}. In the Milky Way, CRs permeating these dense clouds,
lead to an excess of gamma-rays, which emerge from the interaction
between the CRs and the matter.   As a result, GMCs are also regarded as  CR
barometers  \cite{aharonian01}.

The transport of CRs within GMCs is a difficult problem, since GMCs are also complex, evolving
objects which exhibit density fluctuations on all observable scales
\cite{Kainulainen2009}.
In the Milky Way, these clouds have surface densities of a few tens of solar
mass per parsec$^2$ \cite{2020MNRAS.493..351C,2020ApJ...898....3L}. Inside a
cloud, denser, pc-sized regions  called clumps have been observed
\cite{2000prpl.conf...97W}. Several studies suggest that although GMCs are
gravitationally unbound \cite{Dobbs2011}, the dense clumps are pc-sized gravitationally-bound objects whose collapse leads to clustered star formation \cite{2016A&A...586A..68P,2017MNRAS.465..667L}.  The significant density variations can lead to  a non-homogeneous distribution of CRs.  Moreover, the presence of magnetic fields in clouds is one further complication. Although studies have  suggested that the magnetic field strength can be damped in clouds \cite{cesarsky78}, observations have pointed to  moderate magnetic fields ($\approx10 \mu G$) whose strength increases with the gas density \cite{2012ARA&A..50...29C}. The high density and strong magnetic field should form a barrier for CR propagation. This shielding effect might be augmented by the fact that the clumps have centrally condensed density profiles \cite{2007ApJ...665..416K,2018MNRAS.477.4951L,2018MNRAS.474.5588D}. Using Fermi Large Area Telescope (Fermi-LAT) gamma-ray data, we study the shielding of CRs in dense clumps and find evidence where slow propagation is required to explain the gamma-ray deficiency. This slow propagation leads to non-homogeneous CR distributions in clouds.

\section{CR deficit in dense molecular clumps}
We chose Taurus and Perseus clouds to search for the effect of possible shielding of CRs in dense clumps. Located in different layers of the Taurus-Auriga-Perseus complex, the largest local associations of dark nebula \cite{ungerechts87}, Taurus and Perseus are among the nearest GMCs to the Earth, whose
distances are estimated as $100 - 200$ pc and $\sim ~300$ pc, respectively \cite{Dame01}. In the vicinity of these clouds, both low mass and high mass star formations have been observed \cite{walter91}. They are amongst the GMCs with the largest $M/d^2$ values,  where $M$ is the mass and $d$ is the distance to the solar system.  This parameter is crucial for the predicted gamma-ray flux of the GMCs \cite{yang14}, which is proportional to $M/d^2$. High gamma-ray detection significance  allows more accurate spatial and spectral analysis, which is required in this work.   

We use Planck dust opacity maps to trace the gas distributions.  A detailed description can be found in the Methods.  The derived gas column density map is shown is Fig.~\ref{fig:gasdis}, in which we identified six dense clumps with average cubic density higher than $5000\,\rm cm^{-3}$. Of those clumps, five were selected for our investigation (dubbed C1-5), whereas the sixth one is disregarded, as it coincides with the star formation region IC 348. The gamma-ray emission in the direction of IC~348 shows different properties and shall be considered independently (Methods).

We analysed about 12 years of Fermi-LAT gamma-ray data (above $100~\rm MeV$ )
towards the Taurus-Perseus region (see Methods for details on the
analysis). The observed gamma-ray emission in the field of view results from the CR sea interacting with the molecular content (diffuse background) and individual
sources. The former includes the information we want to derive,  where we use a
template obtained from the Planck dust opacity \cite{planck13_11}. The latter is modelled using the fourth source catalog of Fermi-LAT (4FGL) catalogue \cite{4fgl}. A detailed description of  how the diffuse background model was obtained can be found in the Methods. The diffuse gamma-ray
model should be much more complex than the gas distributions traced by the Planck dust opacity, but in the high latitude region  as the one in here, and given the proximity of the GMCs, the total gas in the line of sight is dominated by the GMCs. In such a scenario, the assumption that the gas distribution has a similar spatial distribution to the gamma-ray emission is a good approximation. The usage
of dust opacity as the background template also allows us to separate the dust emission from a different line of sight and derive the CR distribution in different regions. To validate our results, we compared our results with the ones obtained using the galactic diffusion emission model provided by the Fermi collaboration and used in the standard analysis. The consistent existence of the negative residuals showed in  Supplementary Fig.~\ref{fig:resmap_fermi} and~\ref{fig:resmap_dust} demonstrates that the  adopted  model based on the dust distribution is  correct for this region.

\begin{figure}
\centering
\includegraphics[width=0.48\linewidth]{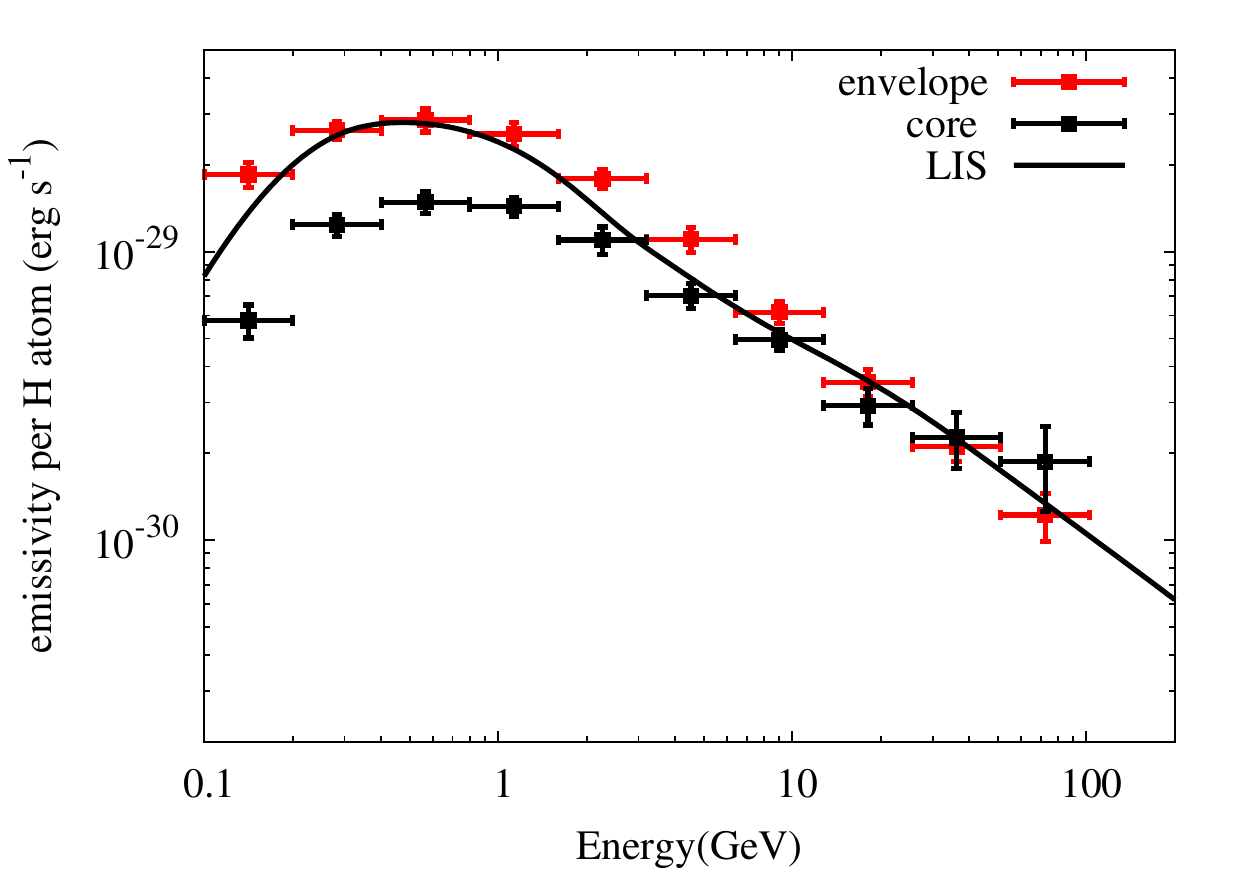}
\includegraphics[width=0.48\linewidth]{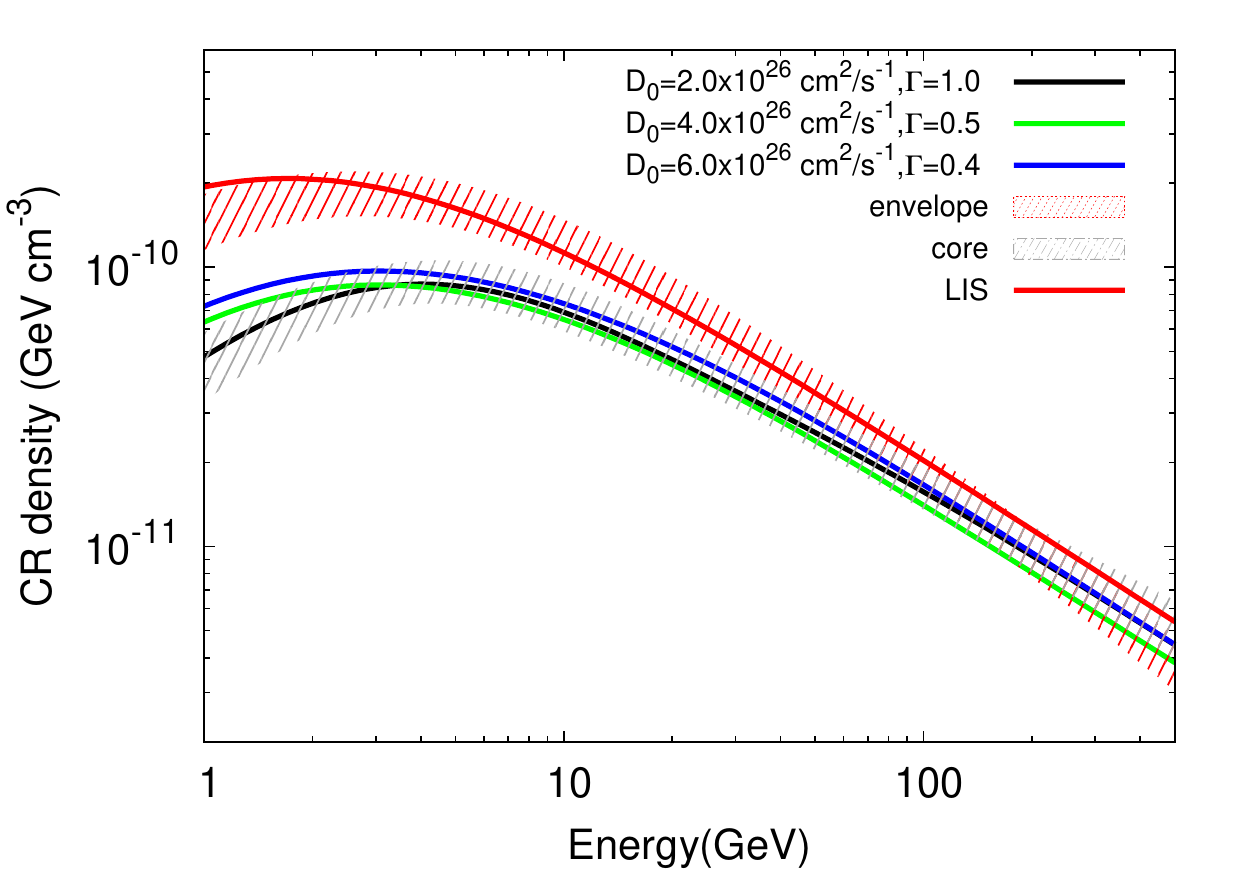}
\caption{ The gamma-ray spectra obtained from Fermi LAT observations and the derived CR energy density.  left panel: The derived gamma-ray spectra of all the clumps normalized to gamma-ray emissivity per H atom both for dense  and diffuse envelop. The x error bars represent the energy bin width, the y error bars are the $1-\sigma$ statistical and systematic errors added in quadrature.
The black curve represent the predicted gamma-ray emissivity assuming the CR spectra is the same as the LIS. right panel: derived CR energy density from the gamma-ray spectra. The red curve is the LIS. The shaded area represent the $1-\sigma$ statistical and systematic errors  for the derived CR density added in quadrature.
 }
\label{fig:sed}
\end{figure}

We separated the Planck dust opacity template into diffuse envelope and dense cores, of which the gas column density is below and above $1.8\times
 10^{22}~\rm cm^{-2}$, respectively. We found such separation can improve the likelihood fitting significantly  when comparing with the fitting without such separation, which implies that the gamma-ray emission from the diffuse envelope and dense cores may have different spectra. This separation also allows us to derive the diffuse spectra of gamma-rays in regions of lower density (envelope) and dense ones  (cores). Then we normalized the gamma-ray emission to the emissivities per H atom,
 which should be proportional to the CR density. The results are shown in the left panel of Fig.~\ref{fig:sed}. We found that the  gamma-ray emissivities derived in the diffuse envelope are consistent with the predicted gamma-ray emissivities, assuming the CR spectra are the same as the local interstellar spectra (LIS). For dense cores, however, the measured gamma-ray emissivities exhibit significant localized deficits at lower energies ($E < 2~\rm GeV$).  
 
One possible explanation of such a deficit of low-energy gamma-rays is that the gas column density of the dense core is overestimated, which lowers the gamma-ray flux consequentially.  This would imply a significantly lower gas-to-dust ratio in dense cores. Indeed, such  a variation of the gas-to-dust ratio has already been observed inside the dense core $\rm \rho~ Oph ~A$ \cite{liseau15}.  However, the very low gas-to-dust ratio is only observed in the very central region of the core ($<0.05 ~\rm pc$), which is far below the resolution of gamma-ray instruments. On the scale of the core of several parsecs, the average gas-to-dust ratio  estimated in ref \cite{liseau15} is consistent with the average value in our Galaxy. Stronger evidence against the interpretation is that in almost all our cases, the deficit appears to be stronger for CRs of lower energies, and an overestimation of the gas column density would affect the normalization of the spectrum but not the shape. The different spectral shapes (as shown in Fig.~\ref{fig:sed}) point rather to different CR spectra at different regions.

We further derived the CR spectra from the observed gamma-ray emissivities. Here, we assumed a smoothed broken power-law spectrum for parent CR protons, that is: 
\begin{equation}
    N_p(E)=A_0 E^{-a_1}(1+(E/E_c)^{a_2-a_1})^{-1},
\end{equation}
where $N_p$ is the proton density, $E$ is the kinetic energy of CR protons, $A_0$ is the normalization factor, $a_1$ and $a_2$ are the spectral indices below and above the cutoff energy $E_c$, respectively. We note that the low-energy data points are not very constraining if $a_1$ is free, thus, we fixed $a_1$ to be 1 in the fitting.  The derived $a_2$ and $E_c$ for diffuse envelope and dense cores are  $a_2= 3.01 \pm 0.06, E_c=2.71 \pm0.65$  and $a_2= 2.87 \pm 0.10, E_c=3.66\pm1.20$, respectively.  The full covariance matrix is used when deriving the CR proton flux, which is shown in the right panel of Fig.~\ref{fig:sed}. The indications of the CR spectra are consistent with those of the gamma-ray emissivities, that is, the CR in the diffuse envelope has nearly identical spectrum as the LIS, while in the dense cores, the CR density is significantly lower below a few GeV, and became consistent with LIS at higher energies. 

The CR deficit at low energies has three possible origins. Firstly, similar to the 'solar modulation' effects in the solar system, this may be due to the young stars formed already inside these dense cores producing strong stellar winds, which prevent the CRs in the ISM to enter the dense cores. Such an effect has also recently been adopted to explain the low CR density in the central molecular zone and other regions\cite{yang14,huang21}.  The effectiveness of this mechanism depends on the overall strength of the stellar winds driven by young stars. The star formation in Perseus and Taurus differs vastly from one region to other\cite{Mercimek2017}, which would result in major differences in the CR deficit from cloud to cloud.  With the current data, we  did not find significant difference in the gamma-ray  spectra for different clumps (Methods),
although given the current statistics, we cannot rule out such a scenario yet. 
Additionally, protostellar outflows are collimated objects, with opening angles ranging from a few degrees for young objects to tens of degrees for evolved objects \cite{2007prpl.conf..245A}. Although these regions might contain a few of these outflows, the outflow cavities are not volume-filling, and their role in CR shielding should be limited.  In view of the above arguments, we would not discuss such a scenario in detail. 

The magnetic mirroring effects can also prevent CRs entering the dense clumps.
As estimated in ref \cite{owen21}, the CR flux can be decreased by about $50\%$ in the clumps compared
with the cores. However, such an effect is also energy independent and thus contradicts to the different spectral shapes observed by gamma-rays (also see the
Methods for details). Thus, the magnetic mirroring cannot account for the entire CR deficit. We note that the effect of the magnetic mirroring is estimated based on the method in \cite{owen21}, in which the time evolution of magnetic fields is neglected. Detailed calculations based on more realistic assumptions is a subject of future studies.

\section{ Slow CR diffusion in clumps} \label{sec:diffusion}
A third explanation that seems to be favoured by our observations is a slower diffusive transport inside the dense cores, which has already been studied in ref  \cite{gabici07}. Under this scenario, the CR deficit is caused by a combined effect of increased proton-proton absorption and slower diffusion caused by an increased magnetic field  \cite{2016MNRAS.459.2432V,2017MNRAS.464.4096L,2017MNRAS.464.4096L,2018MNRAS.474.2167L}. 
Since these factors are independent on star-forming process,
it can explain why CR deficit occurs to all of our clumps. Additionally CRs of higher energies have larger gyroradii, and should diffuse faster \cite{Berezinskii1990book}, therefore the deficit would naturally be stronger for protons of lower energies.

To apply this hypothesis to our observations, we describe the CR transport adopting the equation in ref \cite{gabici07}, that is, CR
transport is dominated by the diffusion equation with energy loss:  
\begin{equation}
    \frac{dN}{dt}=\frac{1}{R^2}\frac{\partial}{\partial R}(D(R,E)R^2\frac{\partial N}{\partial R})+\frac{\partial}{\partial E}(\dot{E} N), 
\label{eq:dif}
\end{equation}
 where $N$ is the space- and energy-dependent particle distribution function of CRs,  $R$ is the distance to the center of the clump, 
$D$  is the diffusion coefficient, and $\dot{E}$ is the energy loss rate of CR protons including both ionization loss and nuclear interaction loss. 
We assume a spherical clump with density profile of
$n(R)=\frac{8000 ~\rm cm^{-3}} {0.04+(R/1~\rm pc)^2} $, and impose the boundary
condition $N(R_{\rm mc})=N_{\rm lis}$, where $R_{\rm mc}$ presents the clump boundary, and
$F_{\rm lis}$ is the LIS CR density  and a reflective (symmetric) boundary condition at $R = 0$. The normalizations are adjusted such that a total of 1800 $M_\odot$ is contained in a region of $\sim 1\;\rm pc$, which is consistent with the  clump parameters in Table~\ref{table:clumps}. Applying the above conditions, and for a stationary solution ($\frac{dN}{dt}=0$),   we can constrain the diffusion coefficient by fitting the derived CR spectrum.   In this work we used a generic form $D(E) = D_0(\frac{E}{1~\rm GeV})^{\Gamma}$,where $D_0$ is the diffusion coefficient 1 GeV
at 1 GeV and $\Gamma$ is the index reflecting the energy dependence.. Since $\Gamma$ is highly unknown, we varied it within the reasonable limits to constrain the value of  $D_0$. Namely,
changing $\Gamma$ within the physically realistic limits from 0 (energy-independent diffusion) to 1 (Bohm-type diffusion) we found that $D_0$ cannot be outside the interval  $2\times 10^{26} - 6\times 10^{26} ~\rm cm^2/s$. Thus, we arrive a robust conclusion that the diffusion coefficient around 1 GeV should be smaller than the one in the ISM by two orders of magnitude.  Three schematic fittings with different $\Gamma$ are shown in the right panel of Fig.~\ref{fig:sed}.  

We conclude that our observations required a slower diffusion characterized by a smaller diffusion coefficient. This stands in contrast to what was
expected in these regions.
According to the prevailing view\cite{cesarsky78}, the
magnetic turbulence should  be damped out in the dense neutral gas
environment, such that the propagation inside these dense cores should be dominated 
by free streaming or advection. The faster transport leads to an effective diffusion coefficient  that is larger than those in the ISM \cite{cesarsky78, morlino15}, which contracts the observational results. However, recent studies reveal that the magnetic field strength increases with the gas density \cite{Crutcher2012}, with, for example, $B\propto \rho^{1/2}$. This can be explained if the energy density of the magnetic field is a fixed fraction ($f_B$) of that of the kinematic motion, which is further tied to the potential energy if the system is virialized \cite{2016MNRAS.459.2432V,2017MNRAS.464.4096L,2018MNRAS.474.2167L}. 
The increased magnetic field in denser regions leads to slower CR diffusion. This scenario can potentially explain the decrease of CRs in dense regions. Another possibility is the magnetic turbulence induced by CR streaming instability, which will also result in a slower diffusion inside dense clouds\cite{dogiel18}. Possible tests to distinguish these two scenarios are the GeV gamma-ray observations on clouds near CR accelerators, such as the molecular clouds interacting with supernova remnants\cite{Jiang2010}, where the CR density are expected to be higher then the nearby GMCs. If the magnetic turbulence is increased mainly due to CR streaming instability, high CR density would imply more severe shielding.  However, such supernova remnant-molecular cloud systems are all quite distant and the angular resolutions of current GeV gamma-ray instruments make such study rather difficult.

The Galactic molecular ISM contains density fluctuations over- all scales. Having established the mechanism of magnetic shielding,  we establish a picture of the effectiveness of magnetic shielding under different conditions. We assume that the clumps are virialised and that the magnetic energy density is comparable to the kinetic energy density. The condition for effective shielding can be derived through $\tau_{\rm diffusion} = \tau_{\rm pp}$, as other effects such as advective transport can be neglected \cite{gabici07}. Here $\tau_{\rm diffusion}$ and $\tau_{\rm pp}$ are the diffusion and cooling time scale, respectively. $\tau_{\rm diffusion} \approx r_{\rm clump}^2/D  $, where $D$ is the diffusion coefficient and $r_{\rm clump}$ is the physical size of the the clump.  Since the inelastic scattering cross section of proton-proton collision is only weakly energy-dependent, $\tau_{\rm pp} \propto n_{\rm clump}^{-1}$, where $n_{\rm clump}$ is the average gas density in the clump. We assume $D\propto B^{-\gamma} E^{\gamma}$, where $\gamma<1$ is needed to  account for
a possible suppression of the diffusion coefficient inside
the turbulent cloud  (in the following we adopt $\gamma = 1/2$) \cite{Berezinskii1990book,Casse2001,Fatuzzo2010}.  

\begin{figure}
    \centering
    \includegraphics[width=0.8\linewidth]{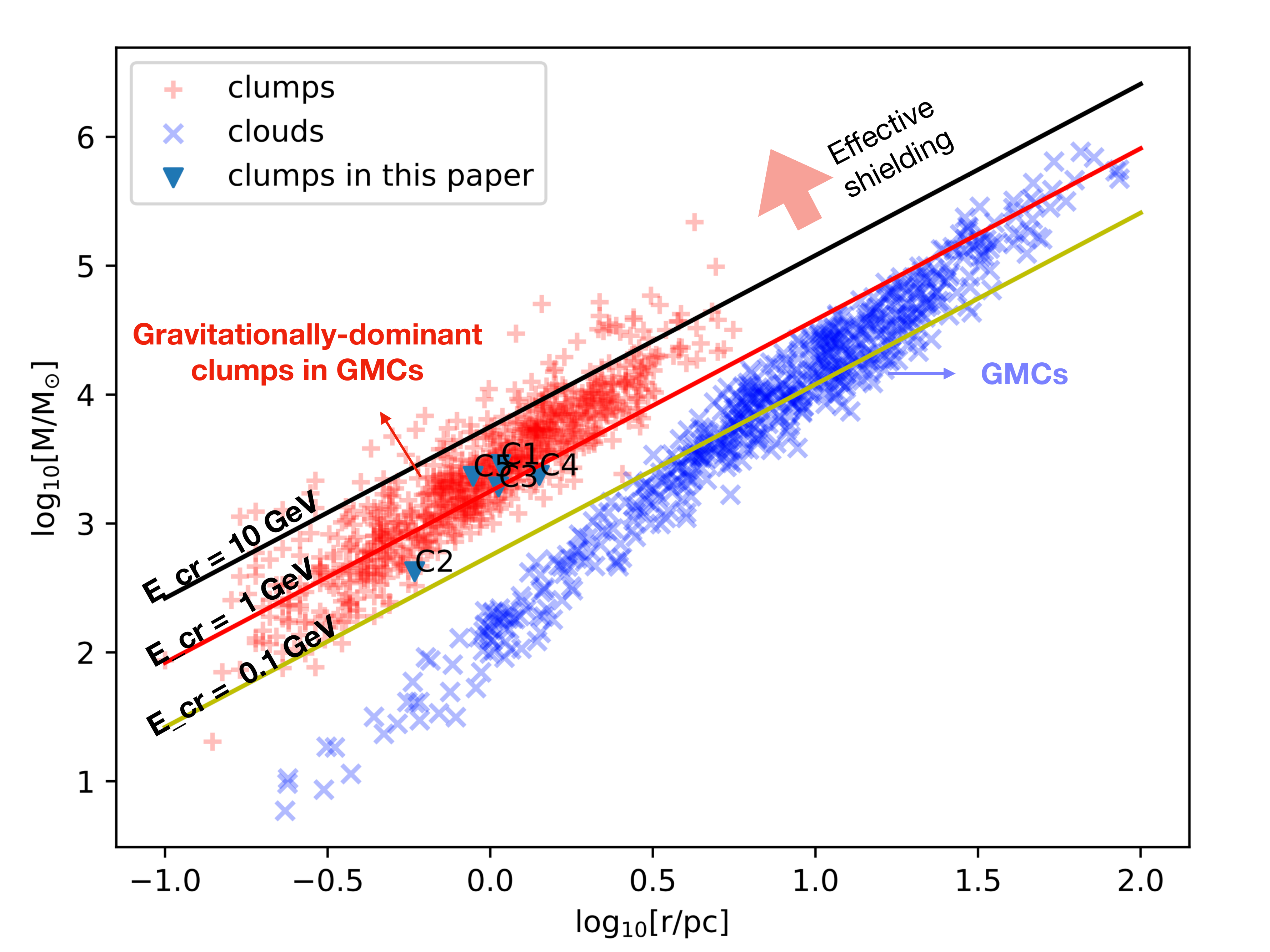}
    \caption{ \label{fig:general:diffusion}  Shielding effect toward
     structures of different masses and sizes. The lines represent different
     thresholds above which shielding becomes important  for CRs of different
     energies. Masses and sizes of  molecular clouds \cite{2020MNRAS.493..351C}
     and star cluster-forming clumps \cite{2014MNRAS.443.1555U} are also
     overlaid.  Shielding is not important on the cloud scale, but becomes significant in dense molecular clumps  where the densities are much
     higher. }
    \label{fig:size}
    \end{figure}

Using Zeeman measurement, observations \cite{Crutcher1999} found out a positive correlation between the gas density and magnetic field strength, a trend that was later refrained by further observations \cite{Crutcher2012}.
This positive correlation can be understood in framework where the magnetic energy density follows the kinetic energy density \cite{2018MNRAS.474.2167L} $B^2 / 8 \pi \approx 1/2\,n_{\rm clump}  \sigma_{\rm v}^2$, where $\sigma_{\rm v}^2$ is the velocity dispersion in the clump. If clumps are self-gravitating and virialized, for example, $\sigma_{\rm v}= (G m_{\rm clump} / r_{\rm clump})^{1/2}$, this scaling relation allows us to estimate the magnetic field strength for clumps of different masses and radii. Thus from $\tau_{\rm diffusion} = \tau_{\rm pp}$ we derive a relation between shielding energy $E_{\rm s}$ and the clump parameters, that is, 
$E_{\rm s}\propto \frac{m_{\rm clump}^3}{r_{\rm clump}^4}$ (where $m_{\rm clump}$ is the clump mass), below which the shielding effect is substantial. 
The details can be found in the section ‘A general model for CR diffusion’ in the Methods. Shielding becomes effective above the threshold of clump mass
\begin{equation}
    ( \frac{m_{\rm clump}}{1800\; M_{\odot}}) = ( \frac{r_{\rm clump}}{1 \;\rm pc})^{4/3} \Big{(}\frac{E_{\rm s} }{1\;\rm GeV } \Big{)}^{1/3}\;.
    \label{eq:mass}
\end{equation}
In Fig.~\ref{fig:general:diffusion}, we plot the
thresholds for effective shielding on the mass-size plane, where locations of molecular clouds \cite{2020MNRAS.493..351C} and clumps \cite{2014MNRAS.443.1555U} are also overlaid. 
The shielding effect is not important on the cloud scales, but becomes notable for the clumps with significantly increased densities.

\section{Conclusions}\

The CRs inside GMCs determine the ionization rate and temperature of the molecular gas and affect the star formation process. The CR density is a crucial parameter in ISM modeling, yet its value is poorly constrained due to our limited understanding of CR propagation. In this Article, for the first time, we find evidence that GeV and sub-GeV CRs, a dominant source of molecular ionization and a major driver of astrochemistry, do not penetrate deep into the dense core of GMCs, leading to localized negative regions (holes) in the gamma-ray residual maps. The effect is more  pronounced at lower CR energies. A direct implication of our results here is the lower ionization rate in these dense clumps. Indeed, the anti-correlation between ionization rates and gas column densities has already been revealed from previous astrochemistry observations\cite{albertsson18}, although not for the exact clumps we investigated in this work.                        

We propose that the CR deficit is caused by slower diffusion in the cloud. 
This slower diffusion of CRs contradicts the prevailing view where  magnetic turbulence is suppressed by neutral-ion damping \cite{cesarsky78}, which enables CRs to penetrate freely into the dense cores. On the contrary, the diffusion should be slower in the dense, star-cluster-forming clumps, where the magnetic field is stronger because the energy density of magnetic field follows the energy density of, for example, turbulent motions \cite{2018MNRAS.474.2167L}. This scenario also explains the energy-dependent CR deficit as indicated by our observations.  

Another possibility is that there is a small region separating the dense clumps from the more diffuse cloud, possibly due to CR self-generated waves \cite{morlino15,ivlev18,dogiel18}, in which the CR diffusion is suppressed. In this case the CR propagation inside clumps can still be dominated by free streaming. The CRs deficit is caused by a barrier at the clump boundary rather than the slow diffusion inside dense clump.

Molecular clouds are objects with significant density fluctuations caused by turbulence and gravity, forming various barriers for CR transport. To describe this process, we propose a self-consistent model to describe the transport of CRs in different regions and find effective shielding occurs at the upper left part of the mass-size plane, above the threshold estimated in  Eq.\ref{eq:mass}. The shielding effect is important in dense,  gravitationally bound regions which are believed to be star-cluster progenitors. In cloud  evolution, gravitational collapse creates high-density regions, where the CR density is reduced and the initial condition of star formation is modified.

\vfill
\clearpage

\section*{Method}
\subsection{Gas distributions and Clump Properties}
\label{sec:Gas}
\begin{figure}
\centering
\includegraphics[scale=0.5]{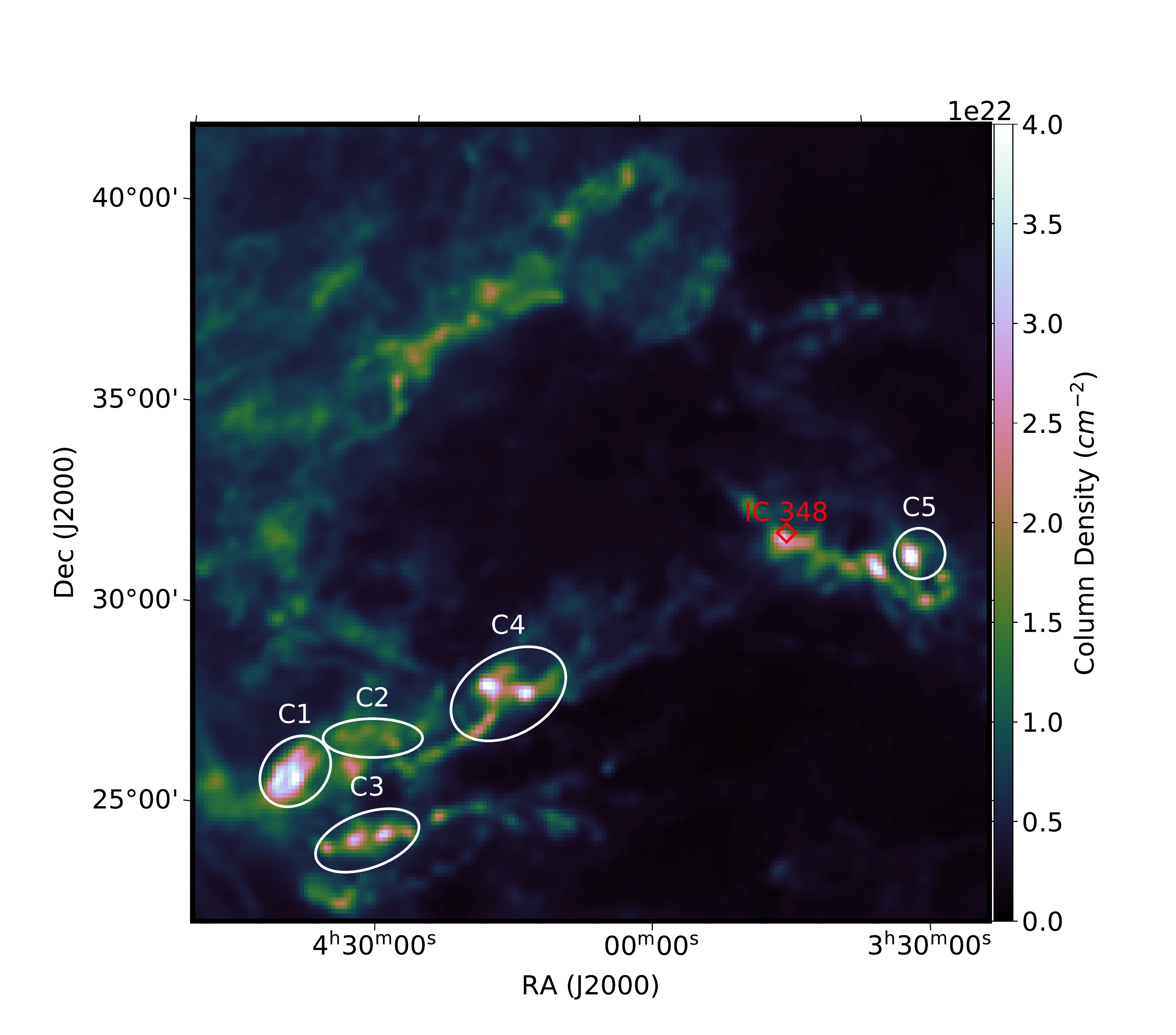}
\caption {
 The total column density (in units of $\rm cm^{-2}$) derived from  the Planck dust opacity map and smoothed with a Gaussian kernel of $0.2^ \circ$. The white ellipses show the position of dense molecular clumps, and the red diamond marks the position of the star cluster IC 348. 
}
\label{fig:gasdis}
\end{figure}

The column density of molecular hydrogen is often estimated using the integrated brightness temperature of the CO emission $W_{\rm CO}$ multiplied by the $\rm H_2/CO$ conversion factor $X_{\rm CO}$. Then, the total gas column density $N_{\rm H}$ can be approximated as $N_{{\rm HI}}+2X_{\rm CO}W_{\rm CO}$,where $N_{\rm HI}$ is the column density of neutral hydrogen (HI). However, such estimation could miss the 'dark gas' that cannot be traced by CO emission.
Thus, for a more robust and reliable result, we use the Planck dust opacity map \cite{planck13_11} to derive the gas distribution, given that the dust opacity is free of the 'dark gas' problem \cite{grenier05}. 
According to Eq.~(4) of ref. \cite{planck11}, the gas column density can be calculated by applying its relation to the dust opacity $\tau_{\rm M}$,  
\begin{equation}
N_{\rm H} = \tau_{\rm M}(\lambda)\left[\left(\frac{\tau_{\rm D}(\lambda)}{N_{\rm H}}\right)^{\rm ref}\right]^{-1}. 
\end{equation}
Here $\tau_{\rm M}$ is a function of the wavelength $\lambda$.
And for the reference dust emissivity measured in low-$N_{\rm H}$ regions, $(\tau_{\rm D}/N_{\rm H})^{\rm ref}$, we adopted the parameters at $353~\rm GHz$ from Table~{3} of ref. \cite{planck13_11} for the calculation.
Note that the dust opacity map contains no velocity information, thus it is impossible to determine the gas content at different distances.  However, the Taurus and Perseus region is located at relative high latitude ($b\approx -20^{\circ}$), and it is natural to assume that the total gas in the line of  sight is at a similar location within the Galaxy.  
The map of gas column density is shown in Fig.~\ref{fig:gasdis}.

The total mass of the cloud in each pixel can be obtained from the expression 
\begin{equation}
M_{\rm H} = m_{\rm H} N_{\rm H} A_{\rm ang} d^{2},
\end{equation}
where $m_{\rm H}$ is the mass of the hydrogen atom, 
$N_{\rm H} $ is the total column density of the hydrogen
atom in each pixel, $A_{\rm ang}$ is the angular area, and $d$ is the distance.  
We chose five clumps in the Taurus and Perseus region, whose mass and locations are shown in the  Table~\ref{table:clumps} and  Fig.~\ref{fig:gasdis}, respectively.

\subsection{Fermi-LAT data}

We selected the Fermi Pass 8 data toward the Taurus and Perseus GMCs from 4 August 2008 (MET 239557417) until 17 January 2020 (MET 600946581)  for the analysis, and chose a $20\deg \times 20\deg$ square region centered at the position of (R.A. = 61.66\deg, Dec. = 32.50\deg)  as the region of interest (ROI). We chose the SOURCE class events with zenith angles less than 90\deg to reduce the contamination from the Earth's albedo. In addition, we applied the filter expression '$\rm (DATA\_QUAL > 0) \&\& (LAT\_CONFIG == 1)$' to select the data of good time intervals that are not affected by spacecraft events. The Fermitools from the Conda distribution available at \url{https://github.com/fermi-lat/Fermitools-conda/}, 
and the P8R3 version of the instrument response functions {\it P8R3\_SOURCE\_V3} are used in the data analysis.

In the background model, we included the sources in the Fermi  eight-year catalog \cite{4fgl} within the  ROI enlarged by 7\deg.
We left the normalizations and spectral indices free for all sources within our ROI. For the diffuse background components, we first applied the standard Galactic diffuse model using the updated template {\it gll\_iem\_v07.fits} and isotropic emission model {\it iso\_P8R3\_SOURCE\_V3\_v1.txt} ( available at \url{https://fermi.gsfc.nasa.gov/ssc/data/access/lat/BackgroundModels.html}) setting their normalization parameters and the index of the Galactic diffuse emission free (such background models are dubbed as Fermi background below).  Since we are investigating the diffuse emission itself,  we also build a background model based on the gas distribution to investigate the spectral characteristics of the diffuse emission in regions of different density.  The later is  built from the gas distribution derived from the Planck dust opacity map \cite{planck13_11} (Planck background model). The galactic diffuse emission basically includes the contributions from the inverse Compton (IC) scattering off soft high energy electrons, as well as  $\pi^0$ decay and bremsstrahlung that take place in the H~{\sc i} and H$_{2}$ regions.
To estimate the background, we calculated the contributions from IC using GALPROP  (\url{http://galprop.stanford.edu/webrun/} ) \cite{galprop}, which uses information regarding CR electrons and the interstellar radiation field.  Isotropic templates related to the CR contamination and extragalactic gamma-ray background were also included in the analysis \cite{3FHL}. And finally we included a background template  generated from dust opacity maps derived by the Planck collaboration  \cite{planck13_11}, where we assumed that gamma-rays trace the spatial distribution of the gas. We chose the size of this background template 10 degrees larger than our ROI, to account for the limited angular resolution in the Fermi-LAT data.  We named such background model as 'dust background'.

\subsection{Spatial structure}
\label{sec:spatial_analy}

To take advantage of the better angular resolution, we used the events above 1 GeV to study the spatial distribution of the gamma-ray emission. We note that there are three point-sources (4FGL J0341.9+3153, 4FGL J0344.2+3203, 4FGL J0346.2+3235) in the vicinity of the star cluster IC348.  To study the excess gamma-ray emission around IC348, we excluded these three 4FGL sources from our background model. We first used the Fermi background model. 
After the likelihood fitting, we subtracted the best-fit diffuse model and all the remaining sources in the ROI, the resulting residual maps are shown in  Supplementary Fig.\ref{fig:resmap_fermi}. We find strong residuals towards the direction of IC 348 region (marked as cyan circle and square), while there are 'negative' residuals in the vicinity of other dense cores.

To study the morphology of the gamma-ray emission located at the position of IC 348, we added a uniform disk on top of the
model used in the likelihood analysis. We then varied the position and size of
the disk to find the best-fit parameters. We compared the overall maximum likelihood of the model with ($L$) (alternative hypothesis) and without ($L_{0}$) (null hypothesis) the uniform disk, and define the test statistics (TS) of the disk model $-2({\rm ln}L_{0}-{\rm ln}L)$ following ref \cite{Lande12}. The best-fit result is a disk centered at (R.A. = $55.98^{\circ}\pm 0.10^{\circ}$,
Dec. = $32.01^{\circ}\pm 0.10^{\circ}$) with $r_{\rm disk}=0.8^{\circ}\pm 0.10^{\circ}$, with a TS value of 140, corresponding to a significance of more than 12 $\sigma$.  We also tested whether this extended emission is composed of several independent point sources. To do this, we recovered the 3 point sources in the likelihood model. Although with more free parameters, the-log(likelihood) function value is larger than that in the 2D disk template case. Thus in the following analysis, we used the
best-fit disk as the spatial template. 

To check whether the negative residuals in the dense cores are from the systematic uncertainty related with the Fermi standard background models, we fixed the normalization of the Galactic diffuse template to be 6\% higher or lower than the best-fit value by hand in the likelihood fitting following the same method used in ref \cite{abdo09}. The derived residual maps are also shown in Supplementary Fig.\ref{fig:resmap_fermi}. We found that, as expected, the fitting is worse than the case when the normalization of Galactic background template are left free, but the 'negative' residual are still significant.

We also used the dust background model to repeat the process. The residual map above 1 GeV is shown in the right panel of Supplementary  Fig.\ref{fig:resmap_dust}. In this case, there are also significant extended excesses observed towards HH211/IC 348. The best-fit result is a disk centered at (R.A. = $55.48^{\circ}\pm 0.10^{\circ}$, Dec. = $31.89^{\circ}\pm 0.10^{\circ}$) with $r_{\rm disk}=0.7^{\circ}\pm 0.1^{\circ}$, with a TS value of 70, corresponding to a significance of more than 8 $\sigma$.

 As shown in Supplementary Fig.\ref{fig:resmap_fermi} and Supplementary Fig.\ref{fig:resmap_dust}, we note that these negative residuals spatially coincide with the densest clumps (marked as C1 to C5). We note that in fact the star cluster IC 348 also coincides with a dense molecular clump.  To test the possible inhomogeneous distribution of CRs inside MCs, we divided the Planck background into dense core and diffuse envelop, with the gas column above and below $1.8\times 10^{22}\rm ~ cm^{-2}$, respectively. The corresponding gas templates are shown in Supplementary Fig.\ref{fig:tmp}.  In the likelihood fitting, we have 5 diffuse components, which are IC and isotropic background, the dense core and diffuse envelop of gas distributions, and the $0.8^{\circ}$ disk template to model the excess in IC 348 regions.  We name such a spatial model as dense + diffuse below. Using the 5 diffuse components described above, together with all the 4FGL sources excluding 4FGL J0341.9+3153, 4FGL J0344.2+3203, and 4FGL J0346.2+3235, the negative residuals are significantly reduced, as shown in the right panel of Supplementary Fig.\ref{fig:resmap_dust}. The improvement of the likelihood fitting are also shown in the -log(likelihood) value as shown in Supplementary Table~\ref{tab:like}. The TS value ( $-2({\rm ln}L_{0}-{\rm ln}L)$) of the dense + diffuse model are about 700, considering the 2 additional free parameters in this model compared with the dust model, the significance is about $26~\sigma$.

To check whether the negative residuals are due to limited angular resolution,  we repeated above analysis procedure using Fermi-LAT data of PSF 3 event class, which has better angular resolution than the SOURCE event class. The derived residual with dust model is shown in the middle panel of Supplementary Fig.\ref{fig:resmap_dust}, which reveals a similar negative residual as in the SOURCE event class. The dense + diffuse model also significantly improve the fitting (see  Supplementary Table~\ref{tab:like}), the TS are about 130.  To further test the possible influence by other diffuse components, we also increase and decrease the normalization of IC and isotropic components by hand in the likelihood fitting, and the results are also shown in Supplementary Table~\ref{tab:like}. The TS of the  dense + diffuse  models are also about 700 in these two cases.   

The derived photon index for the disk templates of IC348 above 1 GeV is $2.2 \pm 0.1$, and the total gamma-ray flux can be estimated as $(1.2 \pm 0.2) \times 10^{-9} \ \rm ph.cm^-2.s^{-1}$ above 1 GeV, which reads $(6.0 \pm 1.2) \times 10^{34} \ \rm erg.s^{-1}$ above 1 GeV, if we assume a distance of 310 pc and a single power law spectrum.

\subsection{Spectral analyses}

\label{sec:spectral_analy}
To further investigate the spectral properties of the GeV emission and the underlying particle spectra, we used the spatial templates mentioned above, and divided the energy range from 0.1 GeV to 100 GeV into 10 logarithmically spaced energy bins and derived the spectral energy distribution (SED) via the maximum likelihood analysis in each energy bin. 
The significance of the signal detection for each energy bin exceeds $2\sigma$.
Besides the  68\% statistical errors for the energy flux densities,  we also varied the normalization of isotropic and IC templates by 6\% artificially to account for the systematic errors associated with the diffuse backgrounds \cite{abdo09}.  The SEDs for the dense core, diffuse envelope and the  IC 348 region are shown in Supplementary Fig.\ref{fig:sedall}. Note that the gamma-ray fluxes are normalized to emissivity per H atom, which is proportional to the parent CR density.  

To further test the possible statistics in the spectral analysis, we used slightly different spatial templates for the dense clumps. Instead of $1.8\times 10^{22}\rm ~ cm^{-2}$, we used the column density $1.3\times 10^{22}\rm ~ cm^{-2}$ and $2.3\times 10^{22}\rm ~ cm^{-2}$  above and below which we divided the dust template to dense core and diffuse envelope. As expected, with a lower column density cut the size of the dense core should be larger and the average density is lower. The derived SEDs for both cases are shown in Supplementary Fig.\ref{fig:sedsys}.  We found the for both cases the gamma-ray emissivities are suppressed compared to LIS in low energy range, and the  suppression is more significant when the column density cut is $2.3\times 10^{22}\rm ~ cm^{-2}$, this is also expected due to a higher density and thus stronger CR shielding.  Finally, we also derived the SEDs using the PSF3 events with  model of Fermi-LAT to take advantage of the best angular resolution. The results are also shown in  Supplementary Fig.\ref{fig:sedsys}, which is in good agreement with the results by using SOURCE data.  

We also derived the gamma-ray SEDs from the individual clumps to see whether there is any difference, the results are shown in Supplementary Fig.\ref{fig:sedclumps}. Due to the limited statistics, we cannot claim significant difference in the derived spectra in different clumps.

\subsection{Deriving the CR proton spectrum}

The CR spectra are derived assuming the CR spectrum have the function form of broken power-law, that is: 
\begin{equation}
    N_p(E)=A_0 E^{-a_1}(1+(E/E_c)^{a_2-a_1})^{-1},
\end{equation}
where $E$ is the kinetic energy of CR protons, $a_1$ and $a_2$ are the spectrum indices below and above the cutoff energy $E_c$, respectively. We note that the low energy data points are not very constraining, thus we fix $a_1$ to be 1. We then fit the observed gamma-ray emissivities using the gamma-ray production cross section parametrised in ref \cite{kafexhiu14}.  The derived $a_2$ and $E_c$ for diffuse envelope and dense cores are  $a_2= 3.01 \pm 0.06, E_c=2.71 \pm0.65$  and $a_2= 2.87 \pm 0.10, E_c=3.66\pm1.20$, respectively.  We then used the full covariance matrix in the fitting to derive the CR proton flux and $1\sigma$ error regions as shown in the shaded area in the right panel of Fig.~\ref{fig:sed}. And from Fig.~\ref{fig:sed} we found the CR fluxes below several GeV are significant lower in the dense cores than those in the diffuse envelope and LIS. 

To see the energy range over which the CRs are effectively shielded, we also plot the predicted gamma-ray emissivities assuming different sharp low energy break $E_c$ in the LIS, i.e., we assume the CR spectra are identical to LIS above $E_c$ and zero below. The results are shown in Supplementary Fig.\ref{fig:specut}, in which we found at least to up to 5 GeV the shielding of CRs should be significant.  

\subsection{Magnetic mirroring} 
In dense clumps, due to the  compression of the magnetic field lines, both the magnetic mirroring and focusing can alter the CR density inside the clumps. We used the method in ref \cite{owen21} to estimate such effects. The scaling factor of CR flux inside the clumps is estimated as $\eta(\chi)=(\chi-\sqrt{\chi^2-\chi})$, where $\chi$ is the enhancement factor of magnetic field inside the clump compared with those in ISM. We note that $\eta(\chi)$ saturated to 0.5 quickly when $\chi >2$. In Supplementary Fig.\ref{fig:mirror} we plot the predicted CR density considering the magnetic mirroring, from which we found that the magnetic mirroring alone can hardly account for the different spectral shapes in clumps and envelopes. We found that such an energy-independent feature can hardly explain the derived CR spectra in clumps and envelopes. A combination of underestimation of mass in clumps as well as the slow diffusion and magnetic mirroring can fit the observations. But the  underestimation of mass from the dust emission reveals a higher gas-to-dust ratio in the clump, which is in contrary with the observations in ref\cite{liseau15}. Another possibility is that the simple evaluation of magnetic mirroring cannot reflect the complexities inside clumps, in which the magnetic fields can be highly turbulent and time-dependent.

\subsection{IC 348}
We then return to the positive residual towards IC 348. The derived gamma-ray SED and derived CR spectrum are shown in Supplementary Fig.\ref{fig:sed_ic348}. We found the extra gamma-ray
emission towards IC 348 can be best modelled with a uniform disk with a radius of
$0.8^{\circ} \pm 0.1^{\circ}$, and the spectrum is significantly harder than LIS. We also
note that the gamma-ray spectrum reveals a 'pion-bump' feature below 1 GeV, thus, we fit this gamma-ray component, assuming they are from the interaction of CRs with the ambient gas. We found the parent proton spectrum described by a power law in kinetic energy is sufficient to provide a satisfactory fit. The derived parent
proton spectrum is also shown in the right panel of Supplementary Fig.\ref{fig:sed_ic348}, and the
best-fit parent proton spectral index is $2.2 \pm 0.1$. We found that this is very similar to the CR spectra derived in young massive clusters (YMC), such as
Cygnus Cocoon \cite{fermi_cygnus}, Westerlund 2 \cite{yang18}, and NGC 3603
\cite{yang17}. Compared with these YMCs, the most massive star in IC 348 is of
the spectral type B5, implying IC 348 should be much less powerful. But IC 348 is also much nearer than those YMCs. Although IC 348 and
NGC 1333 reside in similar environments, that is, the dense cores in Perseus GMC, their spectra are quite different. While in NGC 1333, like in other
dense cores,  the CR spectrum reveals a suppression in low energy, in IC 348, an extra hard CR component is observed.  Ref \cite{luhman16} studied the star
census in both IC 348 and NGC 1333 in detail, and found that the young stars in IC 348 are much more massive than those in NGC 1333. So a tentative explanation would be that a hard component of CRs is accelerated by the massive stars in IC 348, just as the CRs accelerated by YMCs.

\subsection{A general model for CR diffusion} \label{sec:general:diffusion}

We build a comprehensive picture of CR propagation, which is similar to the one presented in \cite{aharonian01}. We assume that the clumps are  virialized where the
velocity dispersion $\sigma_{\rm v} = (G m_{\rm clump} / r_{\rm clump})^{1/2}$, We assume that the energy density of the magnetic field follows the energy density of kinetic motion  $B^2/8\pi \approx 1/2 f_B \rho_{\rm clump}  \sigma_{\rm v}^2$, where $\rho_{\rm clump}$  is the cubic gas density inside the clump.
Assuming clump-scale mean diffusion coefficient of  $D = 1.5 \times 10^{26}  \; {\rm cm}^2 \;{\rm s} ^{-1}  (B / 10 \mu G \times f_B^{-1/2} )^{- 1/2} (E / (1 {\rm GeV}))^{1/2} $,
we find 
\begin{equation}
\begin{split}
n_{\rm H_2, clump} & =   \rho_{\rm clump}/m_{\rm H2}/1.37
    =  m_{\rm clump} / (4/ 3 \pi r_{\rm clump}^3) /m_{\rm H_2} /1.37\\  & \approx 4000 \;{\rm cm}^{-3}(\frac{m_{\rm clump}}{1000\; m_{\odot}}) \Big{(} \frac{r_{\rm clump}}{1 \;\rm pc}  \Big{)}^{-3} \;, 
\end{split}
\end{equation}
where the $n_{\rm H2, clump }$is the cubic density of molecular hydrogen, $m_{\rm H2}$ is the mass of a molecular hydrogen, and the factor 1.37 takes the contribution of heavy elements into account. 
Then the magnetic field strength can be estimated as
\begin{equation}
B = (8 \pi f_B^{-1/2} \rho_{\rm clump}\sigma_{\rm v}^{2})^{1/2} =  96 \mu G f_B^{-1/2} (\frac{m_{\rm clump}}{1000\; m_{\odot}}) \Big{(} \frac{r_{\rm clump}}{1 \;\rm pc}  \Big{)}^{-2}.
\end{equation}

The diffusion time is 
\begin{eqnarray}
    \tau_{\rm diffusion} &=& \frac{ r_{\rm clump}^2}{2D} = \frac{ r_{\rm clump}^2}{  1.5\times 10^{26}  \; {\rm cm}^2 \;{\rm s} ^{-1} (B / 10 \mu G )^{- 1/2} (E /  (1 \;{\rm GeV}))^{1/2}} \nonumber \\
    &=& 3000 \; {\rm yr} \; (\frac{m_{\rm clump}}{1000\; M_{\odot}})^{1/2} \Big{(} \frac{r_{\rm clump}}{1 \;\rm pc}  \Big{)} \Big{(}\frac{E }{1\;\rm GeV } \Big{)}^{-1/2} \;.
\end{eqnarray}
The time for pp collisional absorption is 
\begin{equation}
    \tau_{\rm pp} \approx 2 \times 10^4 {\rm yr}\Big{(} \frac{n}{3000 \;{\rm cm^{-3}}}\Big{)}^{-1} \approx 7500 \;{\rm yr}\; (\frac{m_{\rm clump}}{1000\; M_{\odot}})^{-1} \Big{(} \frac{r_{\rm clump}}{1 \;\rm pc}  \Big{)}^{3} \;.
\end{equation}

and Eq.\ref{eq:mass} can be derived by equating $\tau_{\rm diffusion}$ and $\tau_{\rm pp}$.

\section*{Data Availability}
The Fermi-LAT data used in this work is provided online by the NASA-GSFC Fermi Science Support Center, and can be downloaded from the data serve \url{https://fermi.gsfc.nasa.gov/ssc/data/access/lat/}.The Planck dust opacity map is publicly available in Planck Legacy Archive (\url{http://pla.esac.esa.int/pla/aio/product-action?MAP. MAP_ID=COM_CompMap_ThermalDust-commander_2048_R2.00.fits}) .

\section*{Code Availability}

Fermi-LAT data used in our study were reduced and analysed using the standard FERMITOOLS V1.0.1 software package available from \url{https://github.com/fermi-lat/Fermitools-conda/wiki}.

\section*{Acknowledgments} 
 Rui-zhi Yang is supported by  the NSFC under grants 12041305, 11421303 and  the national youth thousand talents program in China. Guang-Xing Li acknowledges supports from NSFC grant  1227030463, W820301904 and 12033005. Bing Liu acknowledges the support from the NSFC under grant 12103049. Yudong Cui is supported by Ministry of Education of China, and the China Manned Space Project (No. CMS-CSST- 2021-B09).

\section*{Author Contributions}
Rui-zhi Yang, Guang-Xing Li and Bing Liu contributed to the paper in equivalent fractions. Rui-zhi Yang, Bing Liu and Emma de Ona Wilhelmi performed the data analysis, Guang-Xing Li, Rui-zhi Yang, Yu-dong Cui and Felix Aharonian were responsible for the interpretation part, all authors contributed to the manuscript writing.


\section*{Competing Interests}
The authors declare no competing interests. 
\begin{table}
\begin{center}
\caption{Parameters of the five clumps}
\label{table:clumps}
\begin{tabular}{ | c | c c c c c|}
\hline
clumps & center$^1$ & size $^2$  $(r_{\rm maj}, r_{\rm min})$ & distance$^3$ & mass$^4$\,($M_\odot$) & average density$^5$ \,($\mathrm{H\,cm^{-3}}$)\\ 
\hline
C1 &69.0, 25.8  & 1.52, 0.76 & 140\,pc & 2871 & $1.6 \times 10^4$  \\
C2 &67.8, 27.0  & 0.84, 0.41 & 150\,pc & 422 & $1.5 \times 10^4$   \\

C3 &67.87 24.4  & 1.83, 0.61 & 140\,pc & 1923 & $ 1.2 \times 10^4 $  \\ 
C4 &64.1, 28.1 & 1.93, 1.0 & 140\,pc & 2363 & $ 0.6 \times 10^4$   \\ 
C5 &52.2, 31.1  & 1.40, 0.56 & 290\,pc & 2324 & $ 2.4 \times 10^4  $ \\
\hline
\end{tabular}
\end{center}
1. RA and Dec. 
2. Ellipse a and b ($r_{\rm maj}$, $r_{\rm min}$), in parsec. 
3. Distances taken from  \cite{2019ApJ...879..125Z}.
4. The mass is calculated as $m= \int 1.37\; N_{\rm H}\, m_{\rm H}\, {\rm d}s $, where the factor 1.37 takes the contribution of heavy elements into account.
5. The density is computed using $n_{\rm H} = m / (4/3\, \pi\, r_{\rm mean}^3) 1.37^{-1} m_{\rm H}^{-1}$, where $r_{\rm mean} = (r_{\rm maj} r_{\rm min})^{1/2}$.

\end{table}

\newpage
\clearpage

\clearpage

\renewcommand{\figurename}{Supplementary Figure}
\renewcommand{\tablename}{Supplementary Table}

\setcounter{figure}{0}
\setcounter{table}{0}

\begin{table}[h!]
\caption{Fitting results for different models}
\label{tab:like}
\begin{center}
\begin{tabular}{| c|c| c|}
\hline
Model& event class &-log(likelihood) \\
\hline
 dust & SOURCE    & 189284 \\ 
 dust + IC 348 disk & SOURCE    &189259 \\
 dense + diffuse + IC 348 disk &SOURCE   &188910 \\
 dust + IC 348 disk+\%6 bkg$^1$ &SOURCE   & 189271 \\
 dust + IC 348 disk-\%6 bkg$^2$  & SOURCE    &189271 \\
 dense + diffuse + IC 348 disk +\%6 bkg$^1$ &SOURCE   &188917 \\
 dense + diffuse + IC 348 disk -\%6 bkg$^2$&  SOURCE  & 188920\\
 dust + IC 348 disk & PSF3   &156161\\
 dense + diffuse + IC 348 disk & PSF3   & 156095 \\
\hline
\end{tabular}
\end{center}
1. 6\% enhancement of diffuse background.
2. 6\% decrease of diffuse background.
\end{table}

\begin{figure}[ht]
\centering
\includegraphics[scale=0.25]{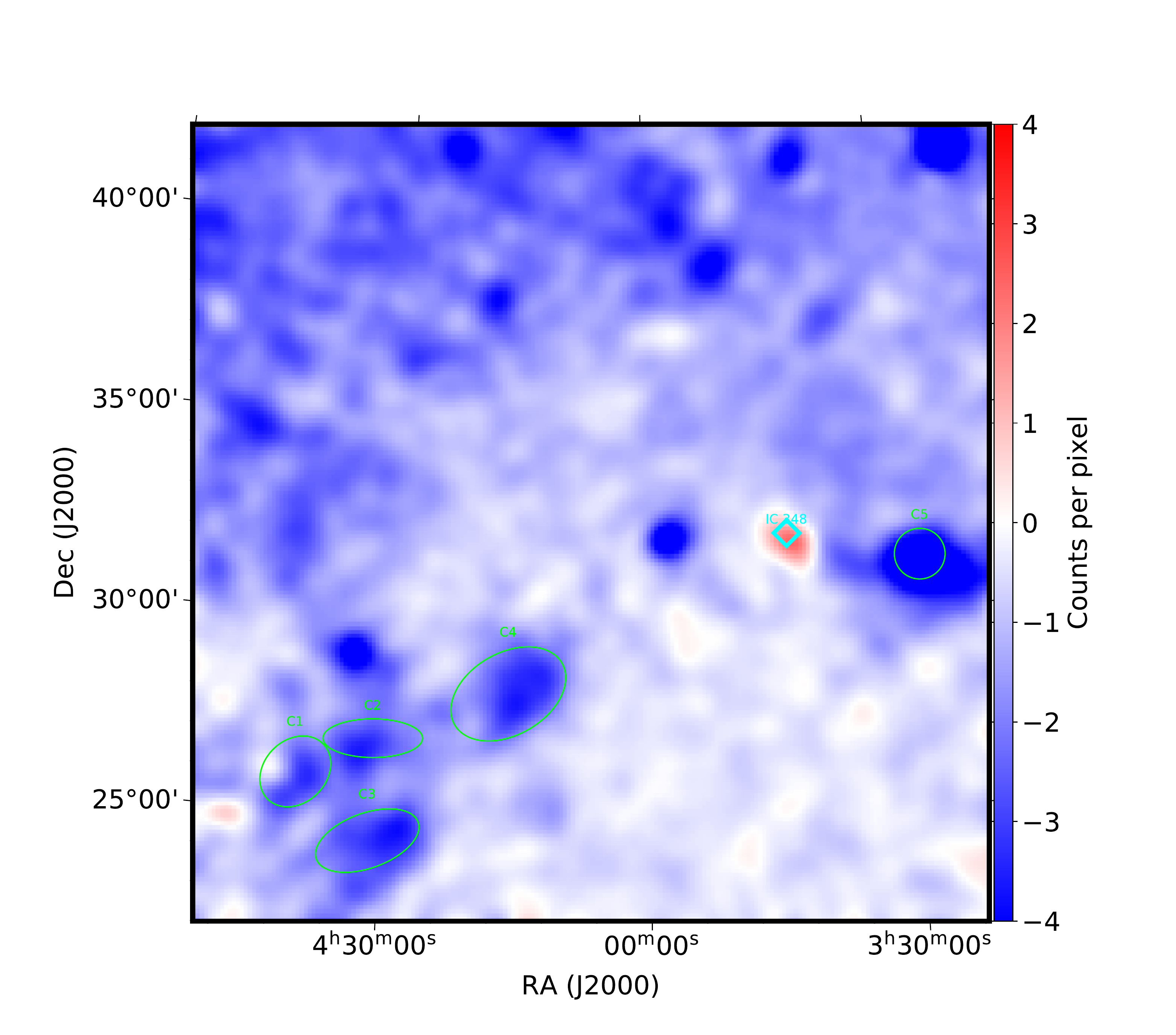}
\includegraphics[scale=0.25]{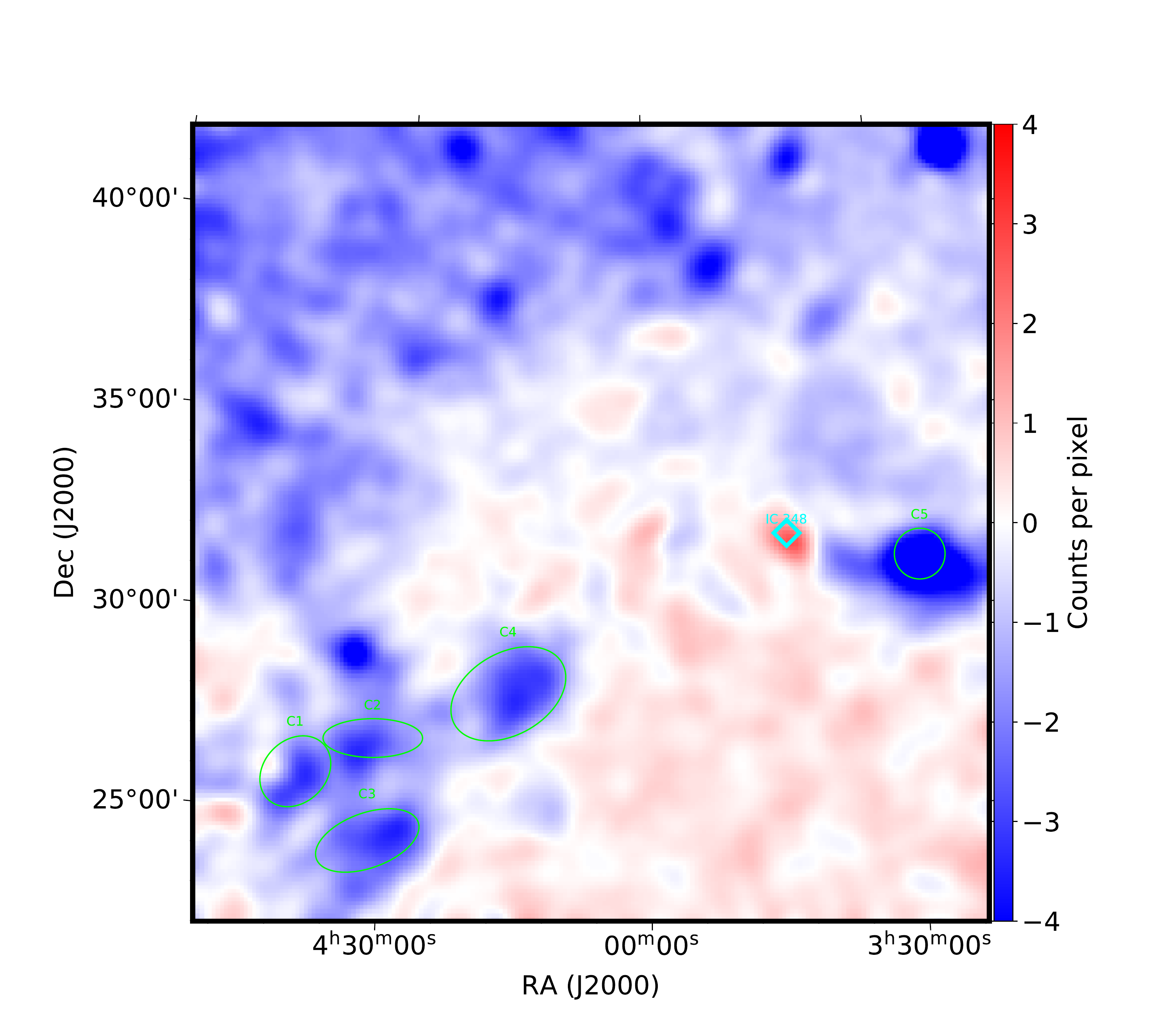}
\includegraphics[scale=0.25]{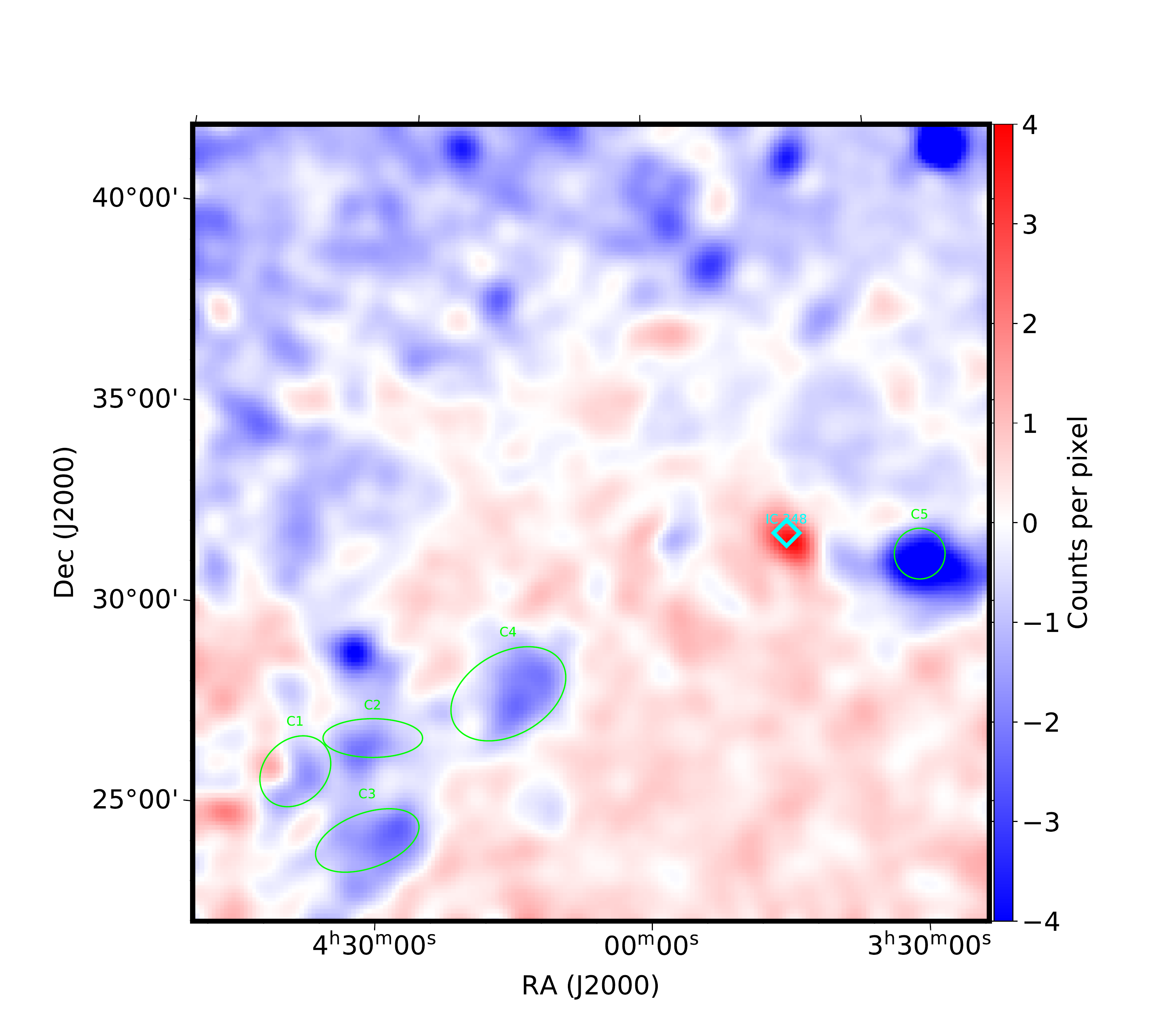}
\caption {The residual counts maps above 1 GeV in the $20\deg \times 20\deg$  region of Taurus and Perseus molecular clouds with pixel size corresponding to $0.1\deg \times 0.1\deg$, smoothed with a Gaussian filter of $0.3^ \circ$. The top left panel shows the result for  the best-fit {Fermi} background, while the top right  and bottom panel show the results by artificially setting the normalization of the Galactic diffuse template to be 6\% higher and lower than the best-fit value,respectively, during the fitting process.
The cyan diamond label the position of IC 348. The green eclipses are the dense cores identified from dust opacity map.  }
\label{fig:resmap_fermi}
\end{figure}

\begin{figure}[ht]
\centering
\includegraphics[scale=0.25]{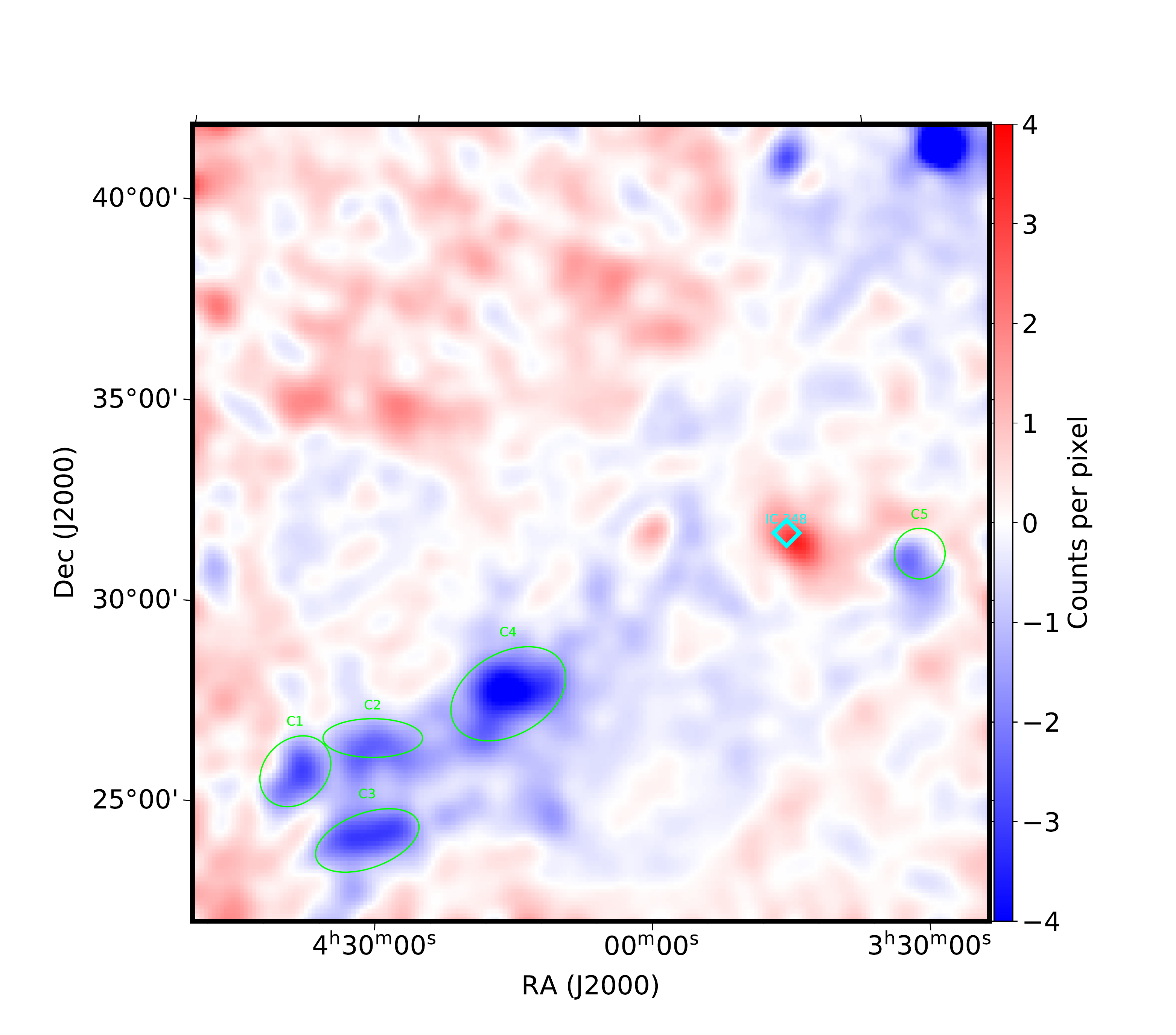}
\includegraphics[scale=0.25]{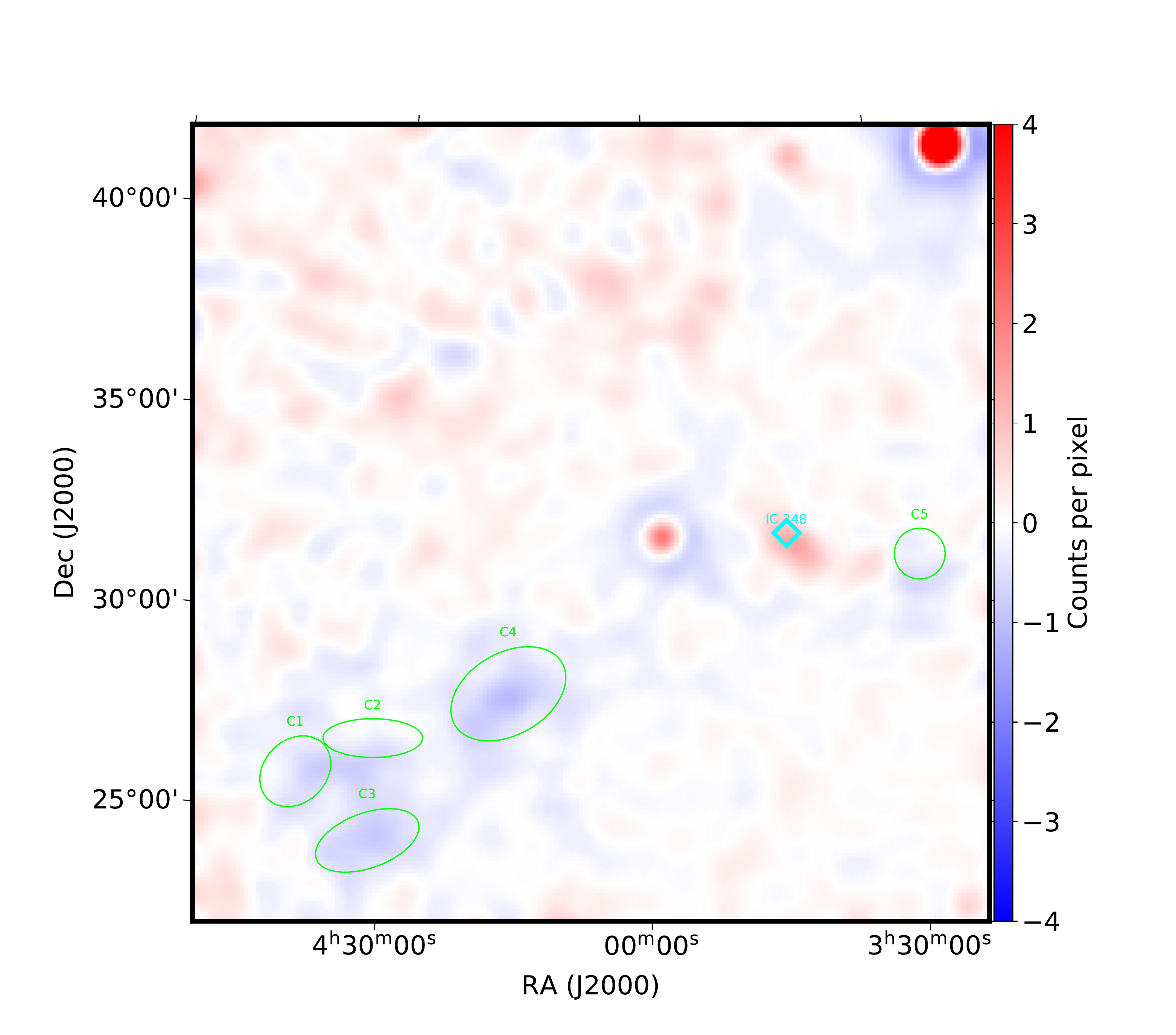}
\includegraphics[scale=0.25]{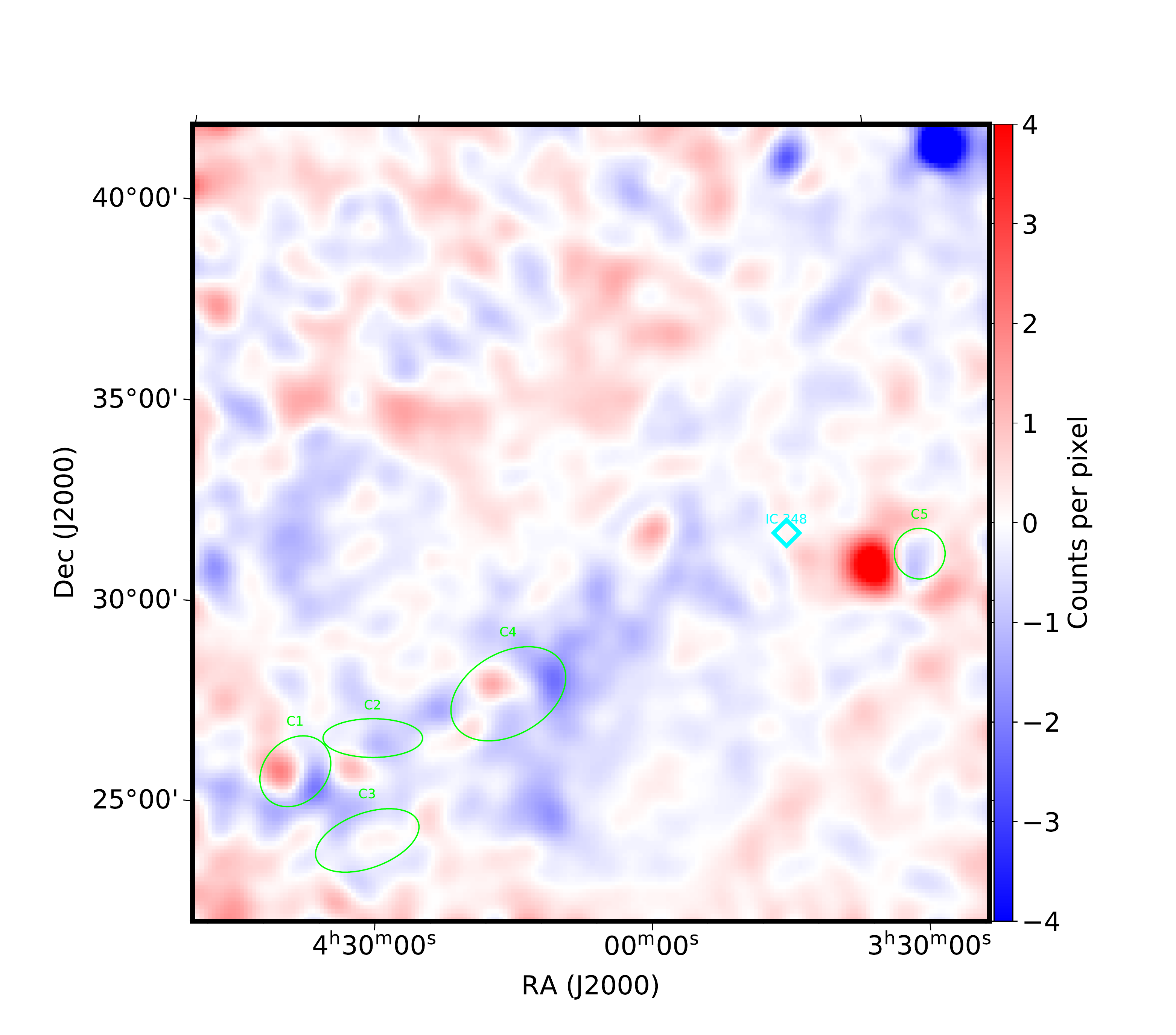}
\caption {The residual counts maps above 1 GeV in the $20\deg \times 20\deg$ region of Taurus and Perseus molecular clouds with pixel size corresponding to $0.1\deg \times 0.1\deg$, smoothed with a Gaussian filter of $0.3^ \circ$. The top left panel shows the result  of dust background with SOURCE event class , while the top right shows the result  of dust background with PSF3 event class, and bottom panel shows the result for {\it dense+diffuse} model.
The cyan diamond label the position of IC 348. The green eclipses are the dense cores identified from dust opacity map.  }
\label{fig:resmap_dust}
\end{figure}

\begin{figure}[ht]
\centering
\includegraphics[scale=0.25]{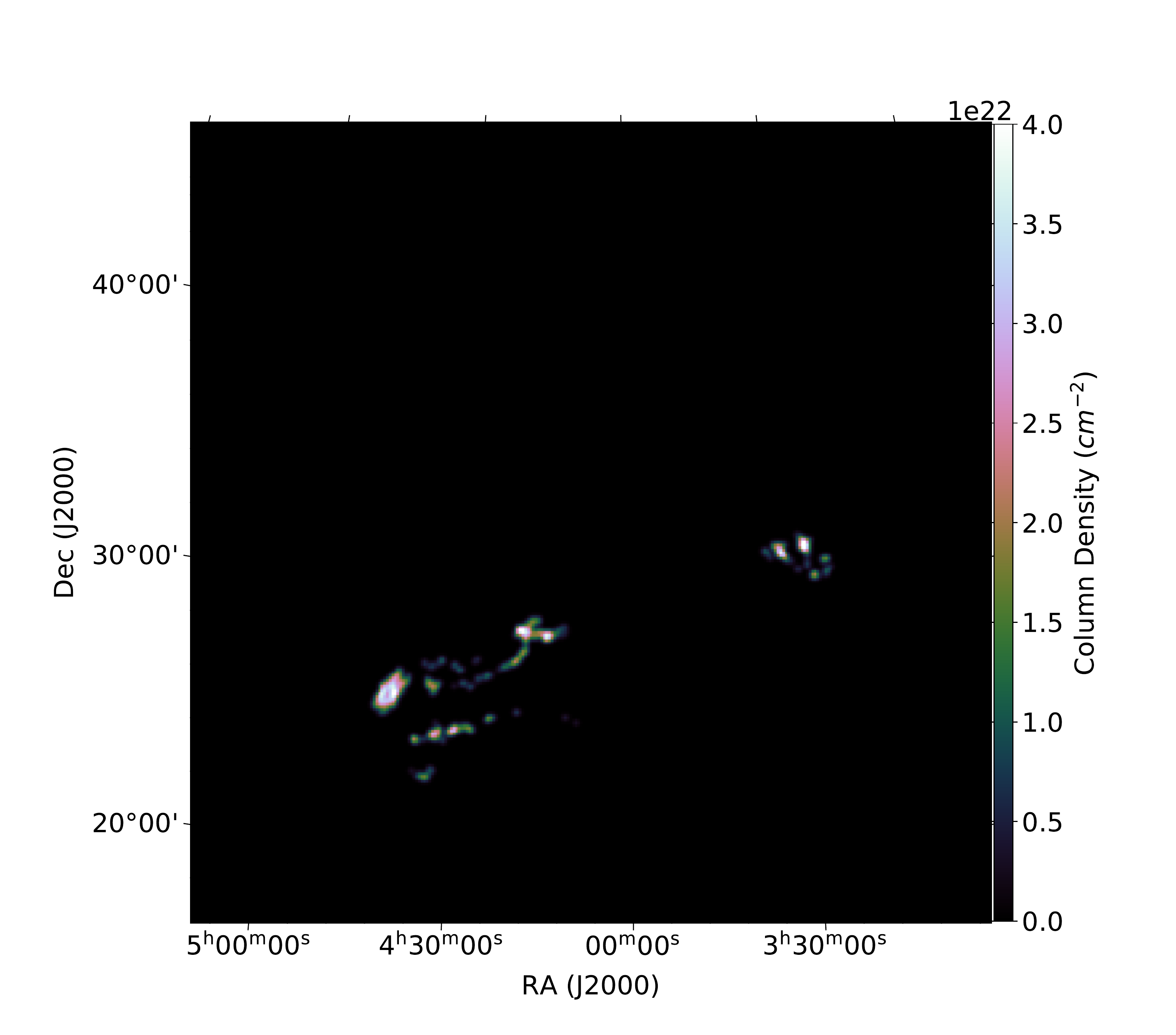}
\includegraphics[scale=0.25]{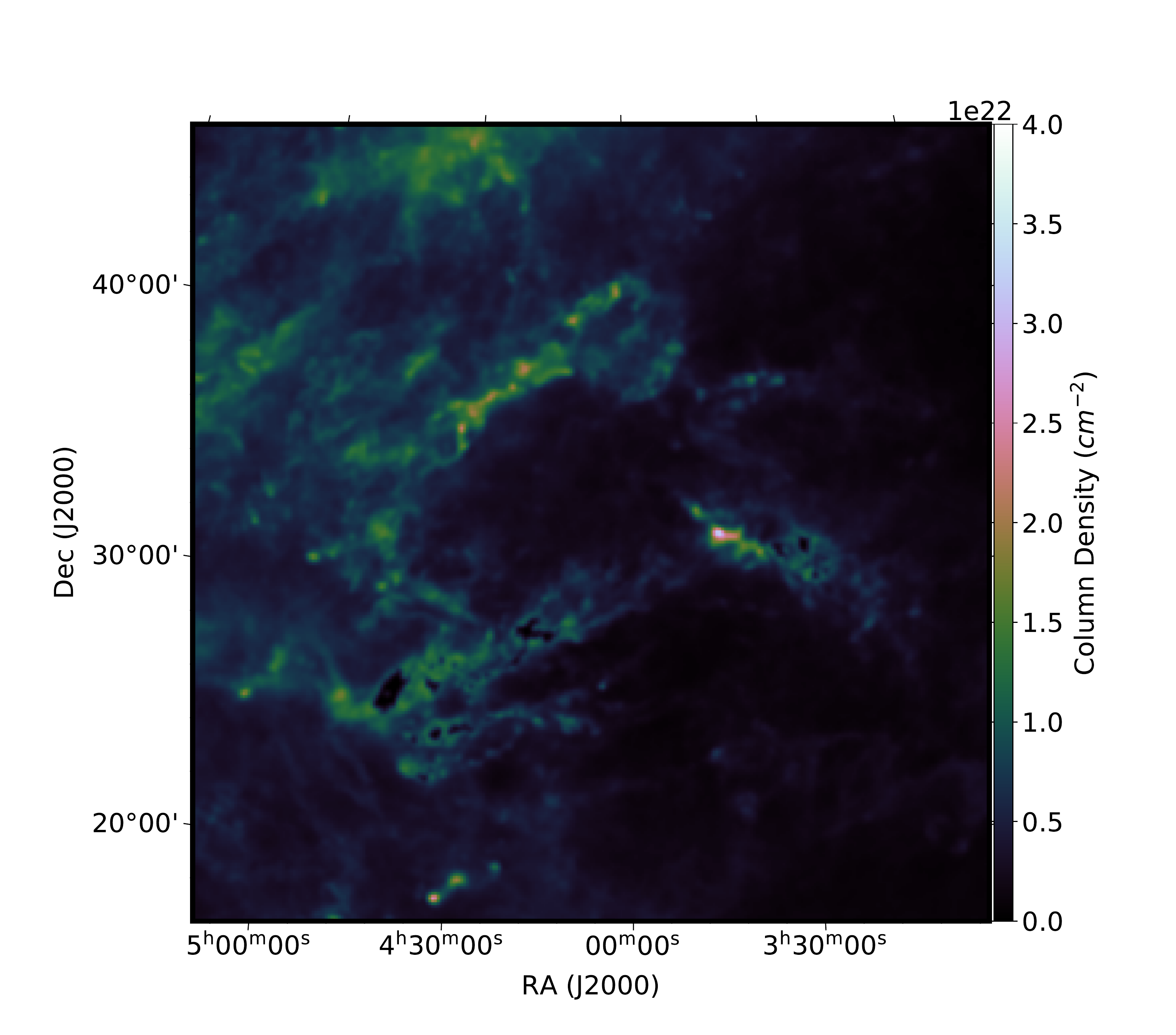}
\caption { The gas column density maps, which are used for gas templates in gamma-ray likelihood fittings,  for dense core (left panel) and diffuse envelope (right panel). }
\label{fig:tmp}
\end{figure}

\begin{figure}[ht]
\centering
\includegraphics[scale=0.7]{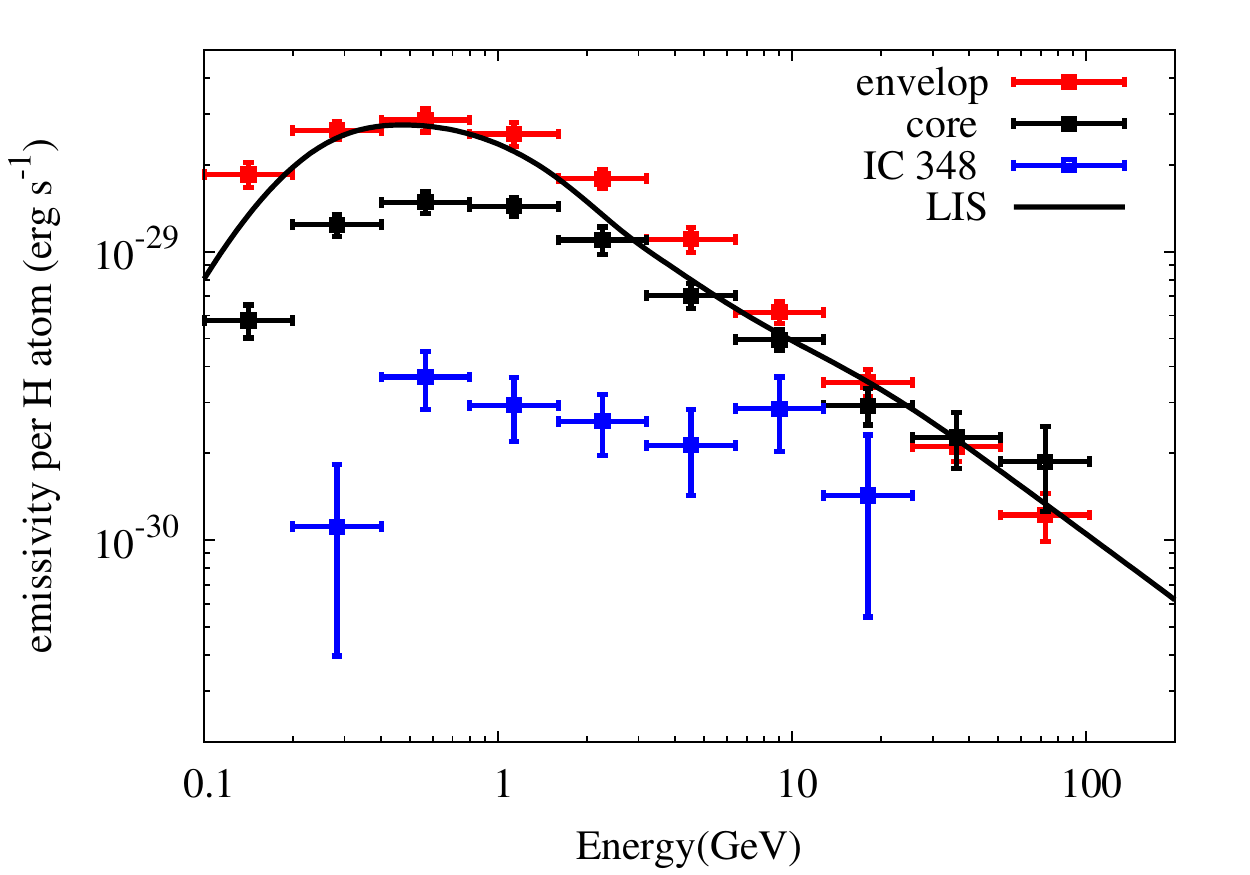}
\caption {
The SEDs of gamma-ray emission of dense core (black), diffuse envelope (red)  and the uniform disk surrounding HH211/IC 348 (blue). The x error bars represent the energy bin width, the y error bars are the $1-\sigma$ statistical and systematic errors added in quadrature.
The gamma-ray spectra have been normalized to gamma-ray emissivity per H atom.  The black curve represents the predicted gamma-ray emissivity assuming the CR spectra is the same as the LIS. 
}. 

\label{fig:sedall}
\end{figure}

\begin{figure}[ht]
\centering
\includegraphics[scale=0.4]{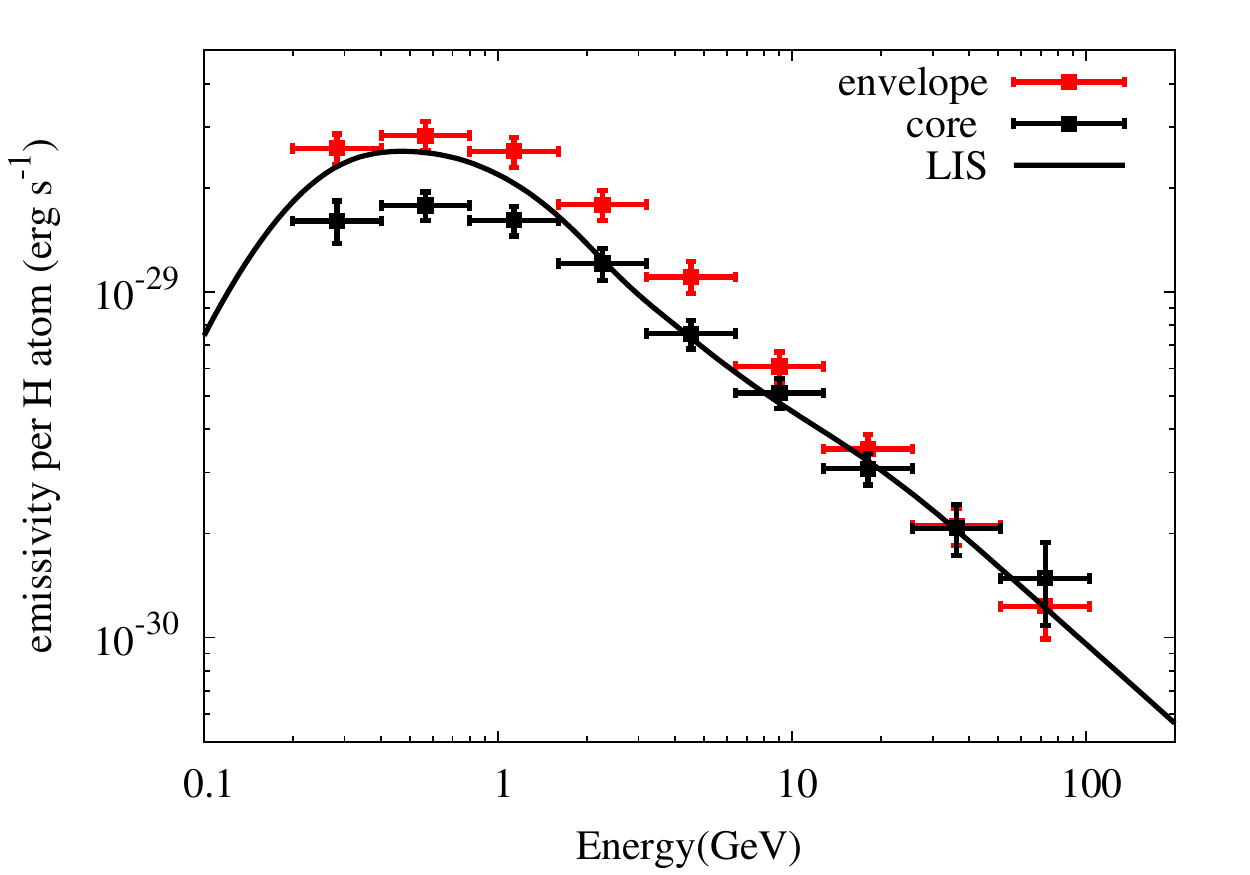}
\includegraphics[scale=0.4]{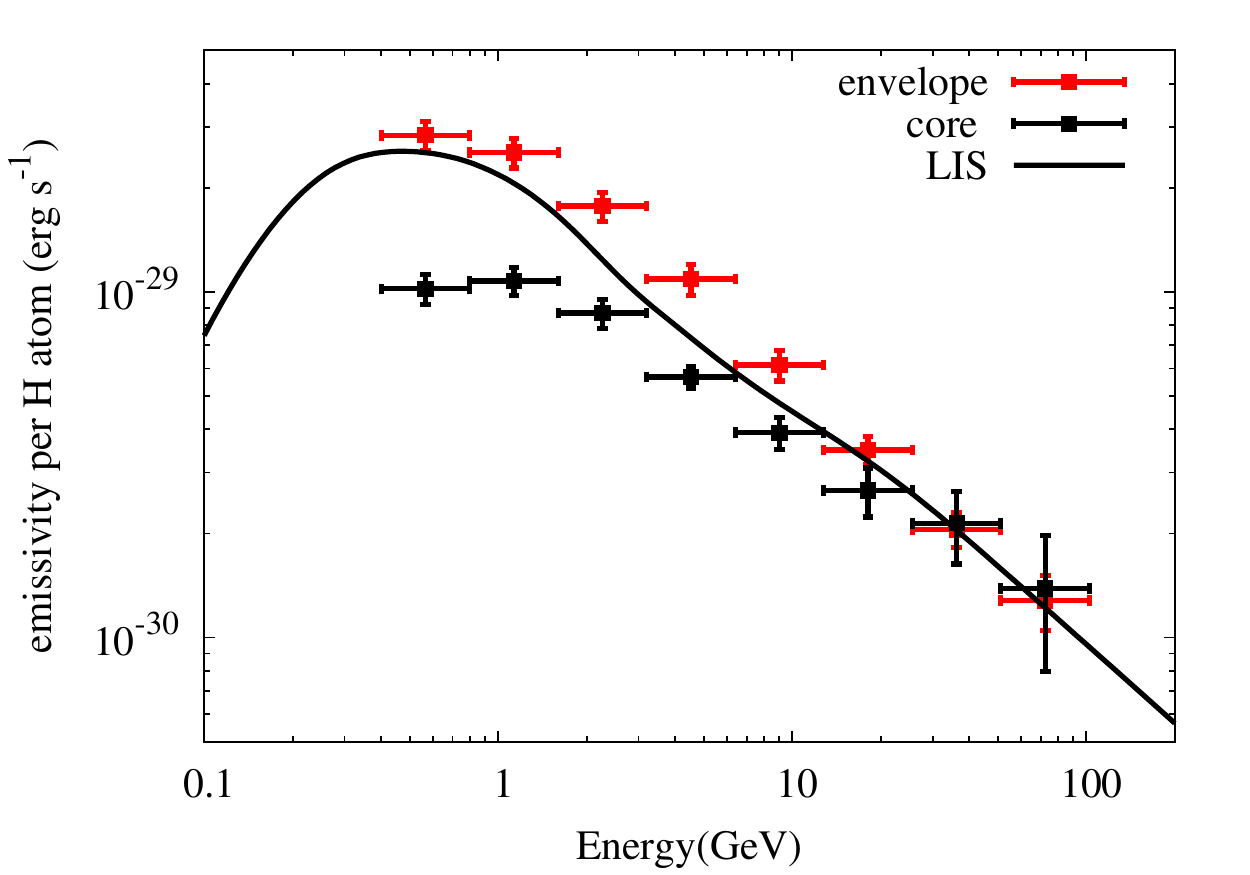}
\includegraphics[scale=0.4]{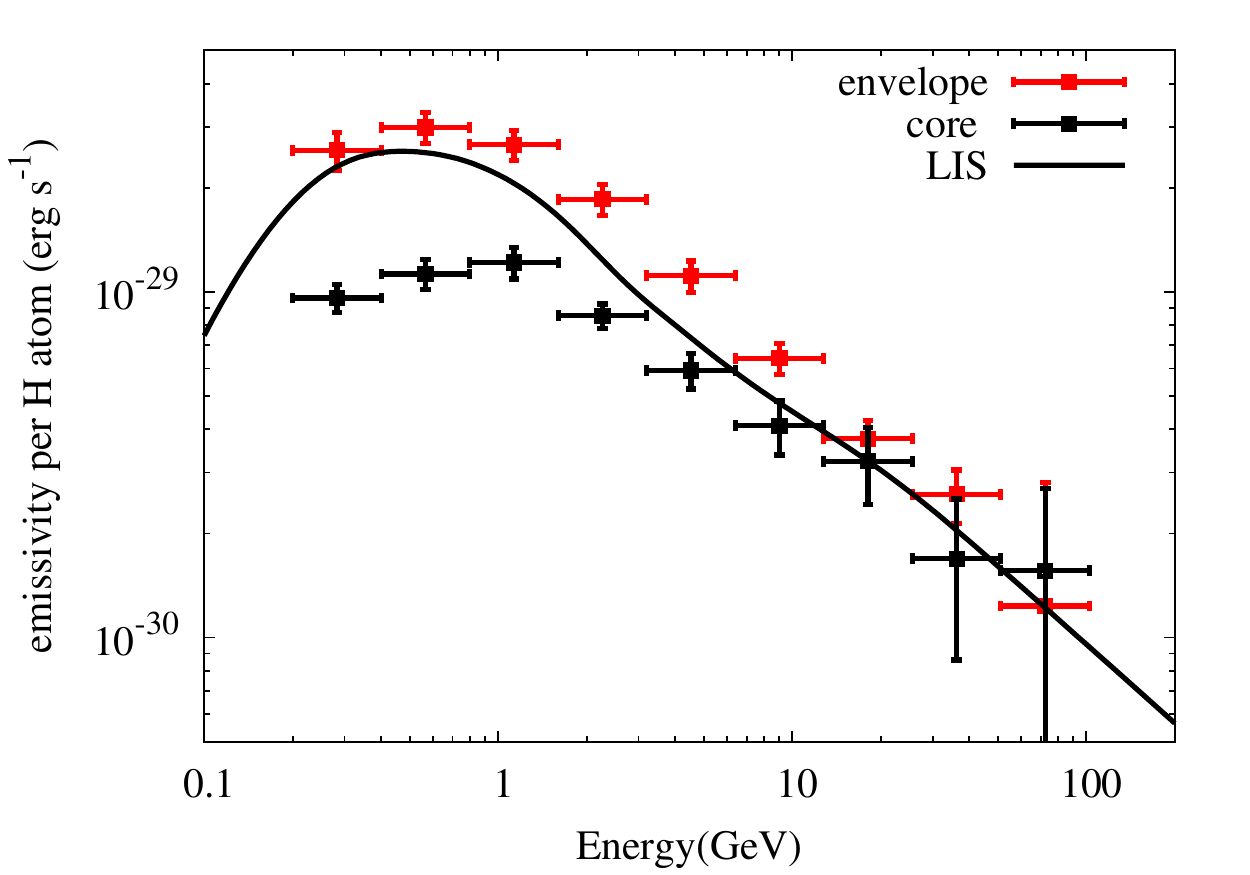}
\caption {
The SEDs of gamma-ray emission of dense core (black), diffuse envelope (red).  The x error bars represent the energy bin width, the y error bars are the $1-\sigma$ statistical and systematic errors added in quadrature.
The gamma-ray spectra have been normalized to gamma-ray emissivity per H atom. The black curve represent the predicted gamma-ray emissivity assuming the CR spectra is the same as the local interstellar spectra (LIS). The left panel shows the results assuming the {\it dense core}  and {\it diffuse envelope} are separated at the the column density of  $1.3\times 10^{22}\rm ~ cm^{-2}$, the middle panel shows the results assuming the {\it dense core}  and {\it diffuse envelope} are separated at the the column density of  $2.3\times 10^{22}\rm ~ cm^{-2}$, and the right panel show the results using the PSF3 datasets with {\it dense+diffuse} model of with the separation is done at $1.8\times 10^{22}\rm ~ cm^{-2}$. }

\label{fig:sedsys}
\end{figure}

\begin{figure}[ht]
\centering
\includegraphics[scale=0.7]{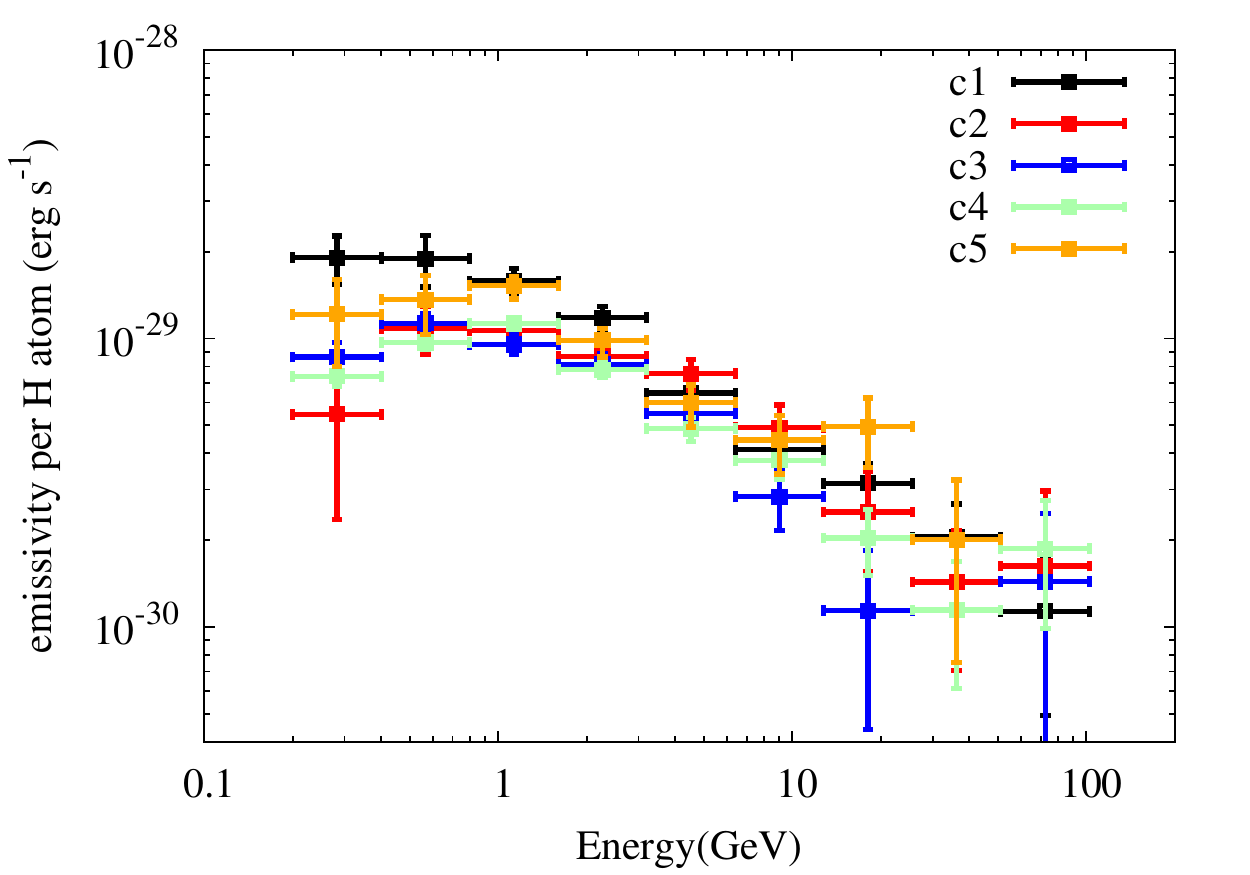}
\caption {
The gamma-ray SEDs of individual clumps (C1-5). The x error bars represent the energy bin width, the y error bars are the $1-\sigma$ statistical and systematic errors added in quadrature.  }
\label{fig:sedclumps}
\end{figure}

\begin{figure}[ht]
\centering
\includegraphics[scale=0.65]{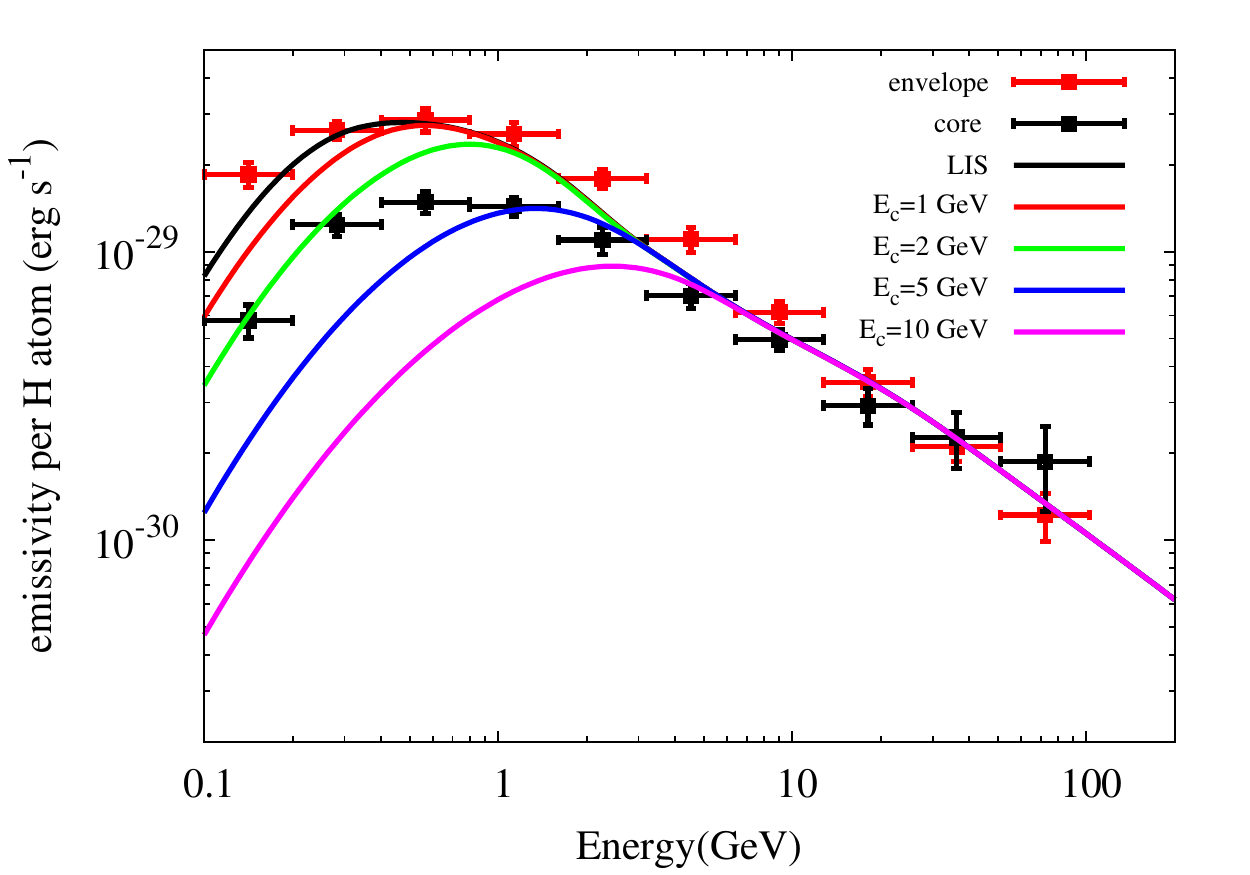}

\caption {
The SED of gamma-ray emission of dense core (black), diffuse envelope (red). 
The gamma-ray spectra has been normalized to gamma-ray emissivity per H atom. The x error bars represent the energy bin width, the y error bars are the $1-\sigma$ statistical and systematic errors added in quadrature. The curves are calculated assuming different sharp low energy cutoff of LIS.   }

\label{fig:specut}
\end{figure}

\begin{figure}
\centering
\includegraphics[width=0.48\linewidth]{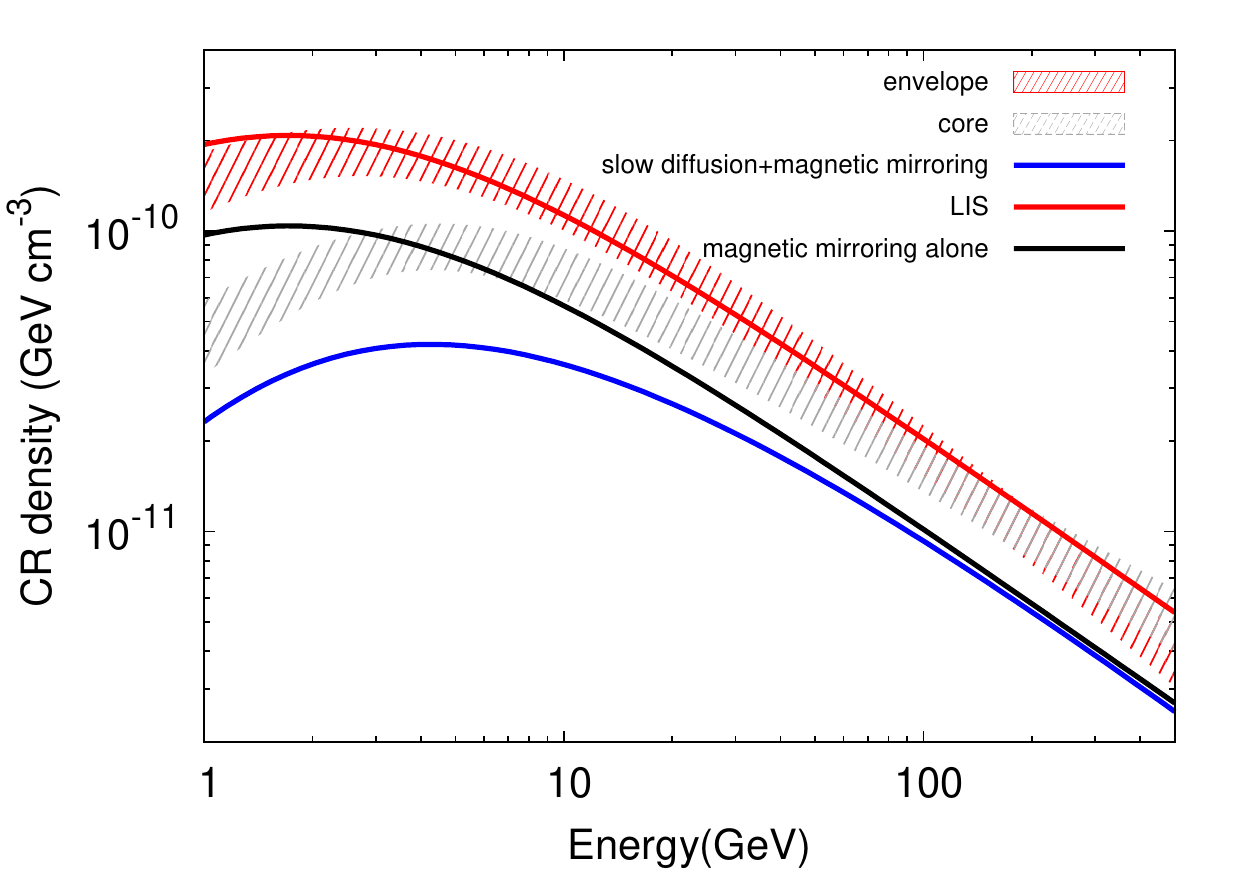}
\caption{ The derived CR spectra and the predicted CR spectrum taking into account magnetic focusing and mirroring. The two curves represent the case with and without the slow diffusion scenario. The shaded area represent the $1-\sigma$ statistical and systematic errors  for the derived CR density added in quadrature.
 }
\label{fig:mirror}
\end{figure}

\begin{figure}
\centering
\includegraphics[width=0.48\linewidth]{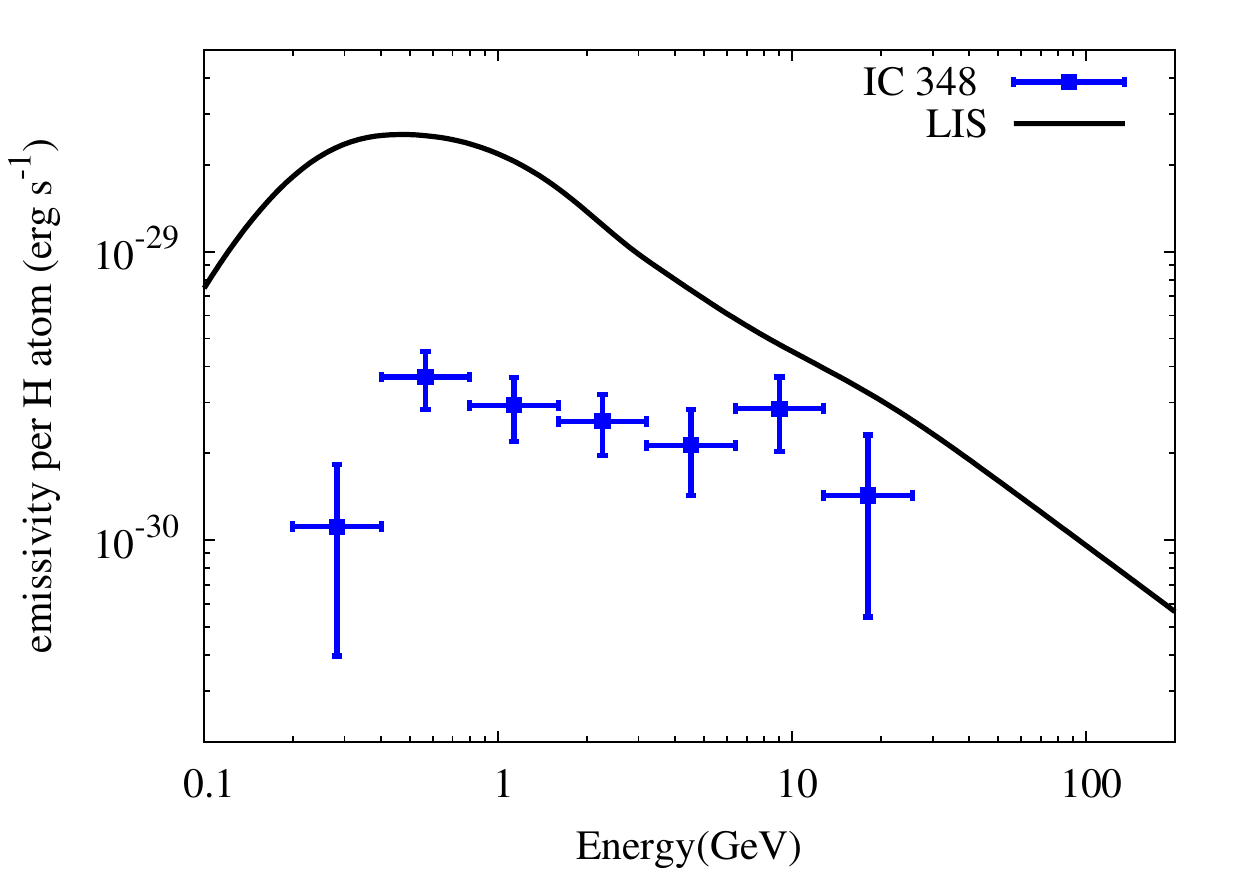}
\includegraphics[width=0.48\linewidth]{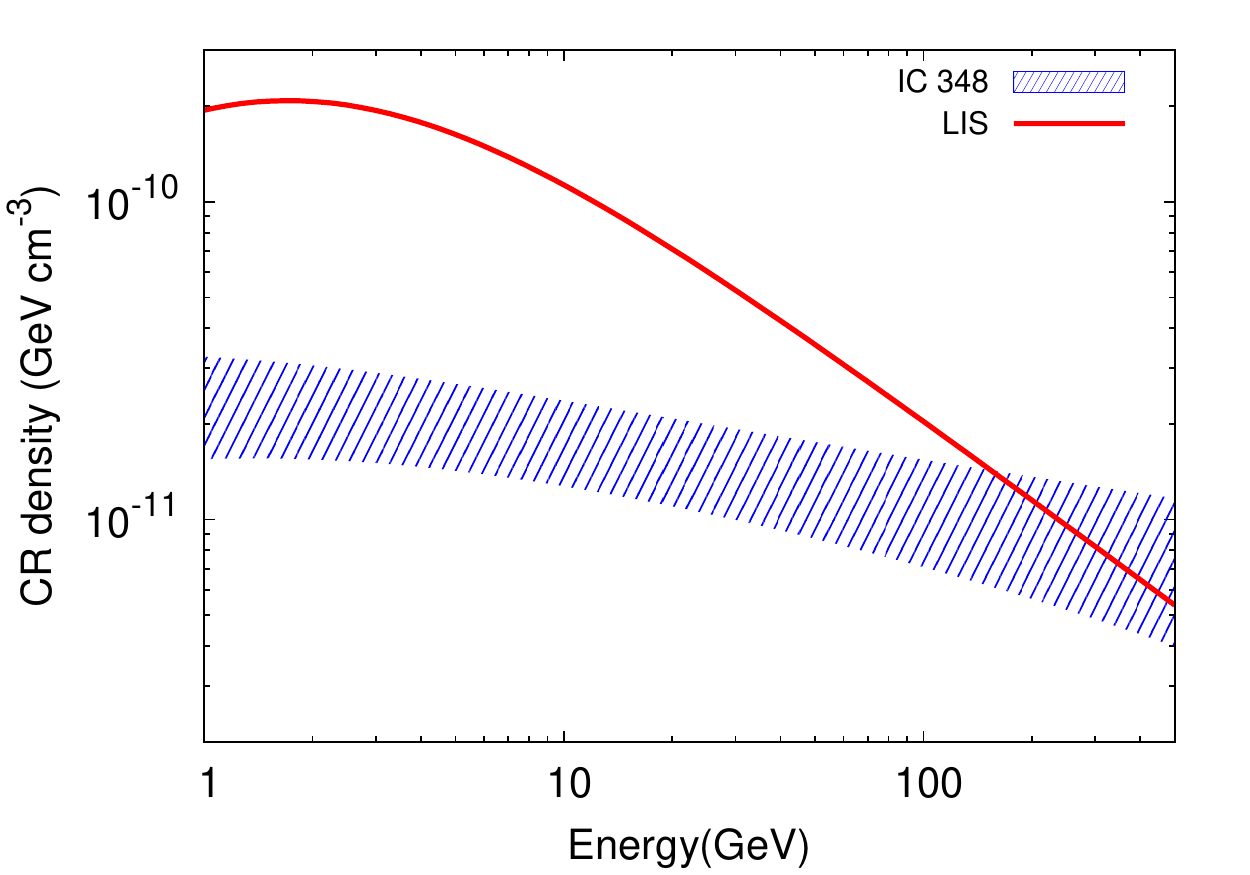}
\caption{ left panel: The derived gamma-ray spectra normalized to gamma-ray emissivity per H atom  for the IC 348 region. The x error bars represent the energy bin width, the y error bars are the $1-\sigma$ statistical and systematic errors added in quadrature. The black curve  represents the predicted gamma-ray emissivity assuming the CR spectra is the same as the LIS. right panel: derived CR energy density from the gamma-ray spectra. The black curve is the LIS.  The shaded area represent the $1-\sigma$ statistical and systematic errors for the derived CR density added in quadrature.
 }
\label{fig:sed_ic348}
\end{figure}

\end{document}